\documentclass{aa}  
\usepackage{natbib}
\usepackage{booktabs}
\usepackage{blindtext}
\usepackage{hyperref}

\bibpunct{(}{)}{;}{a}{}{,} 

\usepackage{graphicx}
\usepackage{txfonts}
\begin{document}

   \title{A study of the electrostatic properties of the interiors of low-mass stars: Possible implications for the observed rotational properties}

   \author{A. Brito
          \inst{1,2} 
          \and
          I. Lopes \inst{2}
          }

   \institute{Instituto Superior de Gest\~ao\\ Rua Prof. Reinaldo dos Santos 46 A, 1500-552, Lisboa, Portugal\\
              \email{anabrito@isg.pt}
         \and
              Centro de Astrof\'{\i}sica e Gravita\c c\~ao  - CENTRA, Departamento de F\'{\i}sica, Instituto Superior T\'ecnico \\ IST, Universidade de Lisboa - UL, Av. Rovisco Pais 1, 1049-001 Lisboa, Portugal\\
              \email{ilidio.lopes@tenico.ulisboa.pt}
             }

   \date{}

 
  \abstract
  {In the partially ionized material of stellar interiors, the strongest forces acting on electrons and ions are the Coulomb interactions between charges. The dynamics of the plasma as a whole depend on the magnitudes of the average electrostatic interactions and the average kinetic energies of the particles that constitute the stellar material. An important question is how these interactions of real gases are related to the observable stellar properties. Specifically, the relationships between rotation, magnetic activity, and the thermodynamic properties of stellar interiors are still not well understood. These connections are crucial for understanding and interpreting the abundant observational data provided by space-based missions, such as Kepler/K2 and TESS, and the future data from the PLATO mission.}
   {In this study, we investigate the electrostatic effects within the interiors of low-mass main sequence (MS) stars. Specifically, we introduce a global quantity, a global plasma parameter, which allows us to compare the importance of electrostatic interactions across a range of low-mass theoretical models ($0.7 - 1.4 \, M_\odot$) with varying ages and metallicities. We then correlate the electrostatic properties of the theoretical models with the observable rotational trends on the MS.}
   {We use the open-source 1D stellar evolution code MESA to compute a grid of main-sequence stellar models. Our models span the $\log g - T_{\text{eff}}$ space of a set of 66 Kepler main-sequence stars.
   }
   {%
   We identify a correlation between the prominence of electrostatic effects in stellar interiors and stellar rotation rates. 
   The variations in the magnitude of electrostatic interactions with age and metallicity further suggest that understanding the underlying physics of the collective effects of plasma can clarify key observational trends related to the rotation of low-mass stars on the MS. These results may also advance our understanding of the physics behind the observed weakened magnetic braking in stars.}
   {}

   \keywords{Stars: low-mass -- Stars:interiors -- Stars: rotation -- Stars: activity -- Stars: evolution
               }
        \titlerunning{Electrostatic properties of stellar interiors}
        \authorrunning{Ana Brito \& Ilídio Lopes}
        \maketitle
%

\section{Introduction}\label{sec:1}

\begin{figure}
        \centering
        \includegraphics[width=9cm]{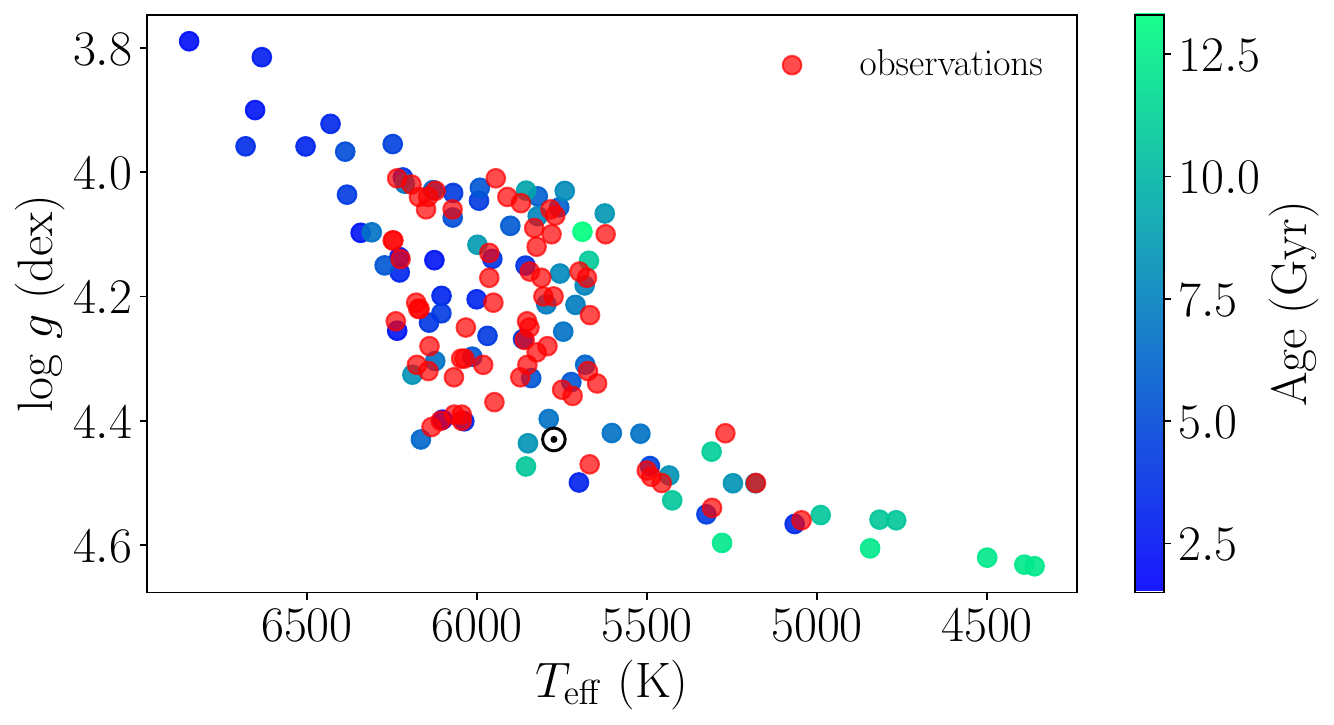}
        \includegraphics[width=9cm]{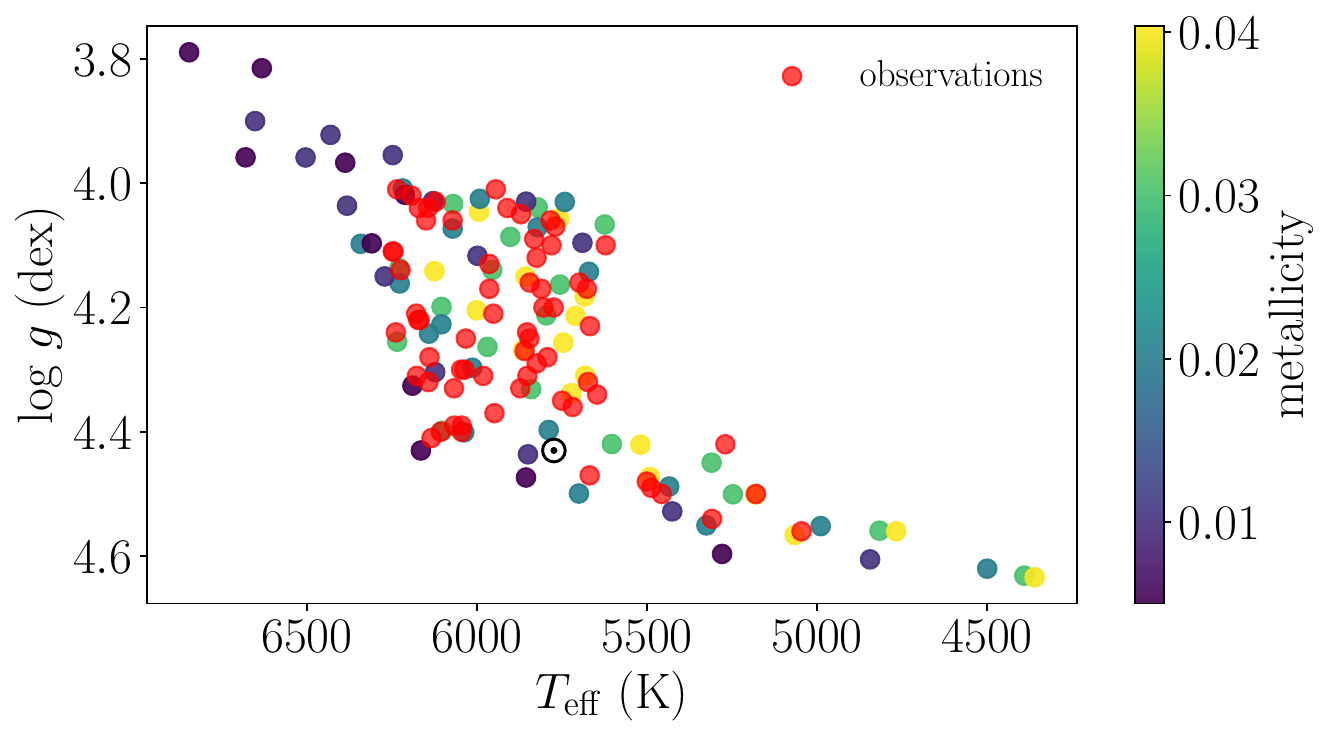}
        \caption{Kiel diagram of the sample of 66 Kepler stars that serve as a basis for the theoretical models. These stars are represented in red. The theoretical data points, representing stellar model values, are colored according to age in the top panel and to metallicity in the bottom panel. The location of the Sun is also represented with its usual symbol for reference.}
        \label{fig1}
\end{figure}

It is well understood that various types of plasma exist in nature, each characterized by significant variations in temperature and electron density. Within the innermost layers of low-mass stars, we commonly encounter what is generally referred to as a hot and dense plasma.  \citet{2016ippc.book.....C} describe plasma as a quasi-neutral
gas of charged and neutral particles that exhibits collective behavior. In this definition there are two key concepts. Quasi-neutrality refers to the fact that the ionic density is not exactly equal to the electronic density ($n_i \approx n_e$). Small local imbalances of electric charge are not only possible but unavoidable, according the plasma definition. Therefore, the plasma maintains an overall state of neutrality, but is not so neutral that it loses all its interesting and  important electromagnetic properties. Collective behavior implies that the motions of plasma particles depend not only on local conditions but also on the state of the plasma in distant regions. Thus, the collective behavior of plasma is a consequence of the long-range properties of electromagnetic forces.

Two important quantities that classify the properties of all astrophysical plasmas, including those found in stellar interiors, are the Debye length and the plasma coupling parameter  \citep[e.g.,][]{2009pfer.book.....M}. The Debye length is a characteristic length scale over which the Coulomb potential of a charged particle is "screened" by the redistribution of the surrounding charged particles of opposite sign. Hence, as a consequence of this Debye shielding, and for scales substantially larger than the Debye length, such as the stellar radii, a state of overall quasi-neutrality is achieved. 
The second important quantity, the plasma coupling parameter, measures the degree of coupling within the plasma, indicating the strength of Coulomb interactions between its constituent particles \citep[e.g.,][]{2009PhRvE..79a6411P, 2016PhRvE..93d3203S}. More specifically, the plasma coupling parameter is a ratio between two magnitudes: the average energy of electrostatic interactions and the average kinetic energies of the particles. This ratio can serve as an indicator of the dynamics of the plasma as a whole; that is, it is an indicator of the significance of the collective effects of the plasma.

In the initial part of this study, we investigate the properties of critical quantities from the perspective of plasma physics. These quantities enable us to describe the electrostatic effects within the interiors of low-mass stars. Specifically, we examine the properties of the Debye length and the energy density of electrostatic interactions throughout the stellar interiors. Special attention is given to the plasma coupling parameter, along with some related quantities such as electron number density, electron degeneracy parameter, and mean molecular weight. For the first time, we explore how the internal Coulomb interactions between charged particles compare across a large set of theoretical models of main sequence (MS) low-mass stars. We computed the internal profiles of the plasma coupling parameter for a group of models with masses ranging from $0.7$ to $1.4 \, M_\odot$. For each model, we define a global parameter that allows us to classify the star from the perspective of Coulomb interactions. Additionally, we analyze how this global parameter varies with mass, age, and metallicity.
%


\begin{table*}
        \caption{Basic statistics (theoretical models vs observational sample).}              
        \label{tab1}      
        \centering                                      
        \begin{tabular}{c c c c c c}          
                \hline\hline                        
                & Mass ($M_\odot$) & $T_{\text{eff}} (K)$ & Age (Gyr)  & Z & $\text log \, g $  \\    
                \hline                                   
                Model average / Sample average & 1.08 / 1.08 & 5823 / 5888  & 6.16 / 6.08 & 0.0202 / 0.0120 & 4.24 / 4.20 \\
                Model SD / Sample SD & 0.192 / 0.153 & 514 / 269 & 2.95 / 2.70 & 0.124 / 0.0140 & 0.210 / 0.156 \\
                \hline                                             
        \end{tabular}
        \tablefoot{
                Comparison of the average values and their standard deviation (SD) for stellar parameters such as mass, age, metallicity, effective temperature, and surface gravity between models and observational data.
        }
\end{table*}
  
\begin{figure*}
        \centering
        \includegraphics[width=6cm]{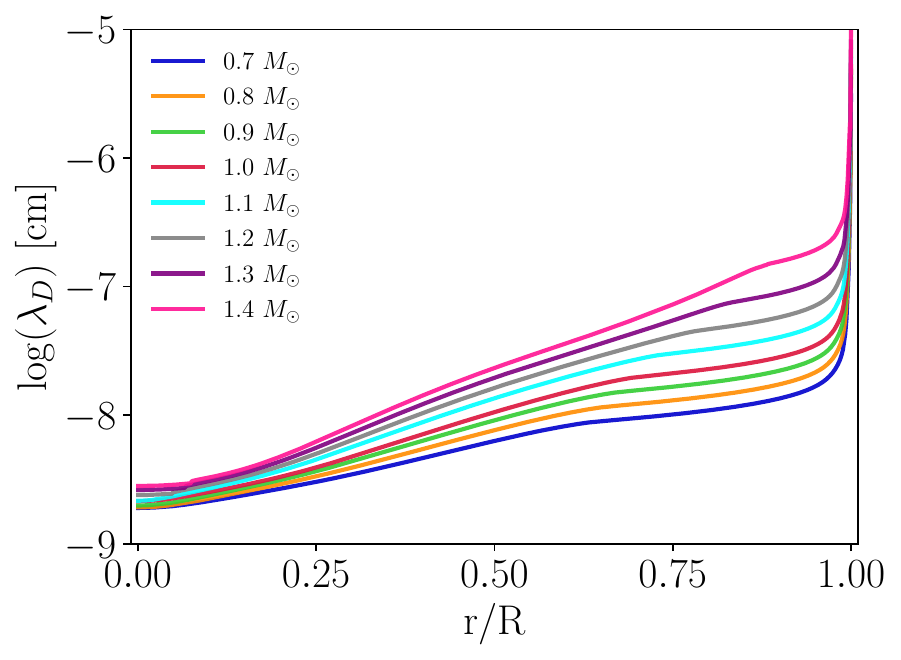}
        \includegraphics[width=6cm]{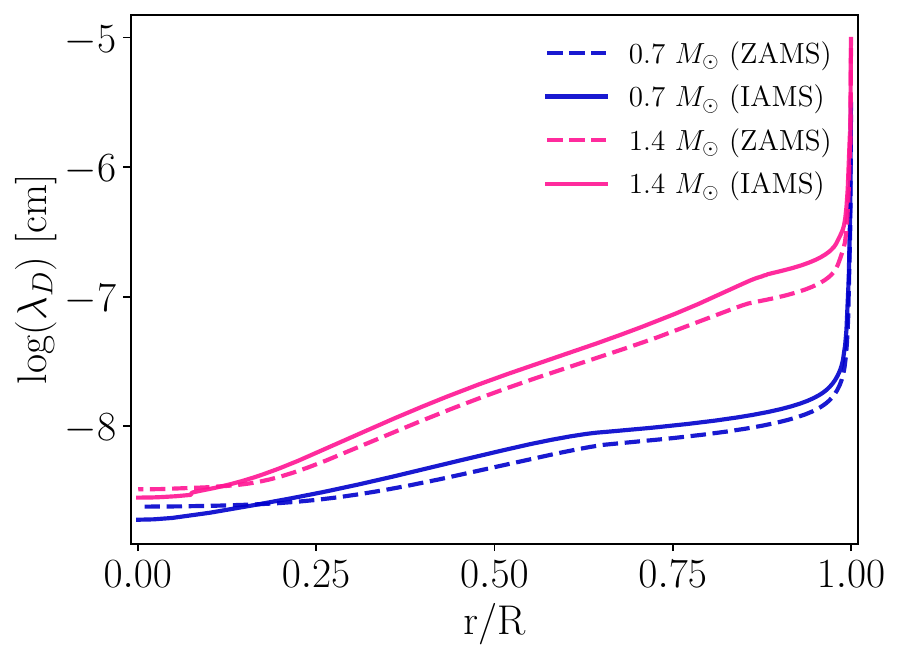}
        \includegraphics[width=6cm]{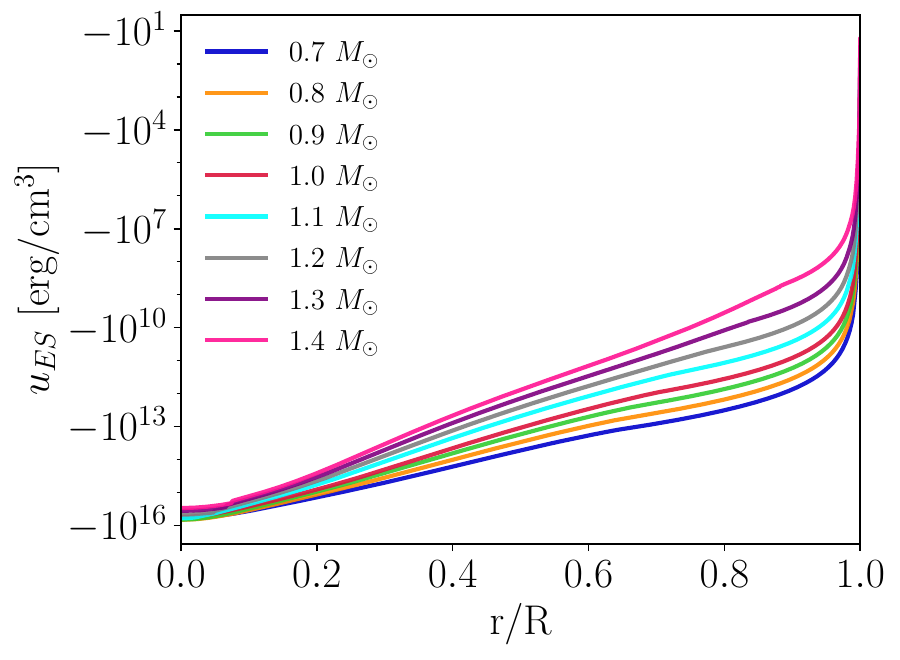}
        \caption{Internal profiles of the Debye length as a function of the fractional radius. All models represented were computed with $Z = 0.02$. Left panel: Changes in the Debye length profiles for models with masses in the range $0.7-1.4 \, M_{\odot}$. Central panel: Debye length profiles for models with $0.7$ and $1.4 \, M_{\odot}$ at two different stages of evolution: at the ZAMS and at midlife. Right panel: Internal profiles of the total energy density of electrostatic interactions as a function of the fractional radius. All the models in this figure are at the IAMS, meaning they are in the middle (intermediate age) of their evolutionary path on the main sequence.}
        \label{fig2}
\end{figure*}

Stellar magnetic fields are crucial for understanding stellar structure and evolution, stellar oscillations, and the magnetic activity and space weather around the stars. In particular, for low-mass MS stars, the dynamo-generated magnetic fields are thought to be tied to the formation of stellar winds, which carry away angular momentum, supporting the stellar spin-down observed in these stars \citep[e.g.,][]{1962AnAp...25...18S, 1967ApJ...148..217W, 1988ApJ...333..236K, 2014MNRAS.441.2361V}.  
Magnetic activity manifestations, such as chromospheric or coronal emissions, have long since revealed strong correlations between stellar rotation and magnetic activity \citep[e.g.,][]{1991LNP...380..353H, 1995A&A...294..515H, 2007ApJ...657..486B, 2014A&A...572A..34G, 2014MNRAS.444.3517M, 2014ApJ...790L..23D, 2016A&A...590A.133O}.  
Generally, it is found that rapid rotators exhibit higher levels of magnetic activity than slow rotators  \citep[e.g.,][]{1967ApJ...150..551K}. The transition from slow to fast rotators is believed to be related to the efficiency of magnetic braking, which occurs due to angular momentum loss through magnetized stellar winds. Low-mass stars below the Kraft break have deep convective envelopes capable of hosting dynamos that produce efficient magnetized winds, contributing to the observed spin-down in these stars. On the other hand, stars above the Kraft break have shallower convective envelopes that cannot host efficient dynamos, resulting in weaker magnetized winds and rapid rotation \citep[e.g.,][]{2013ApJ...776...67V}.

Collectively, the observational evidence of the activity--rotation relationship   obtained over recent decades has allowed us to establish empirical relations between the stars' rotation periods and their ages. \citet{1972ApJ...171..565S} was the first to show that the equatorial rotational velocities of low-mass stars are proportional to the inverse of the square root of stellar age ($v_{\text{eq}} \propto t^{-0.5}$). Therefore, rotation periods can be used to estimate the ages of stars that spin down on the MS, which is the core principle of gyrochronology, a method used to estimate the ages of isolated stars  \citep[e.g.,][]{2003ApJ...586..464B, 2008ApJ...687.1264M, 2010ApJ...721..675B, 2014ApJ...780..159E}.

  It is now clear that observations related to the stellar rotation of MS stars reveal that rotation is an intricate function of mass, age, and metallicity. Hence, the study of the internal structure and internal thermodynamics of stars is fundamental for understanding all the available observational traits. We are interested in studying the microphysics of stellar interiors, which can be linked to the rotational patterns observed on the MS for low-mass stars. Therefore, in the second part of this work, we study the relationship between the electrostatic properties of the interiors of low-mass MS stellar models and the observed rotational properties of stars with similar characteristics.  In particular, given the plasma-like characteristics of stellar interiors and the importance of Lorentz forces in the transport of angular momentum \citep[e.g.,][]{2022MNRAS.517.3392Z},  we relate the properties of the herein introduced global plasma parameter with the different observed rotational behaviors of low-mass stars on the MS.  We find that the electrostatic properties of stellar interiors correlate with the observed MS rotational trends for low-mass stars.

The first part of this work, consisting of sections \ref{sec:2} and \ref{sec:3}, focuses on the study and description of electrostatic effects within stellar interiors. Section \ref{sec:2} describes the stellar models and the modeling process. In section \ref{sec:3}, we examine the main properties of the Debye length, the total energy density of electrostatic interactions, and we define the global plasma coupling parameter that allows us to compare the electrostatic properties of stellar interiors across different stellar models. We also define and examine global values for the electron number density, the electron degeneracy parameter, and the mean molecular weight.

In the second part of the work, we explore the correlations between the global plasma parameter and the observed rotational trends on the MS. Section \ref{sec:4} is dedicated to studying how the global plasma parameter varies with age and metallicity, whereas section \ref{sec:5} compares the theoretical scaling of the global plasma parameter with the observed scaling of stellar rotation rates. Finally, we present our conclusions in section \ref{sec:6}.


\section{Modeling the $\log$ \texorpdfstring{\MakeLowercase{g}}{g} --  $T_{\text{\texorpdfstring{\MakeLowercase{eff}}{eff}}}$ space of Kepler MS stars}\label{sec:2}

\begin{figure*}
        \centering
        \includegraphics[width=6cm]{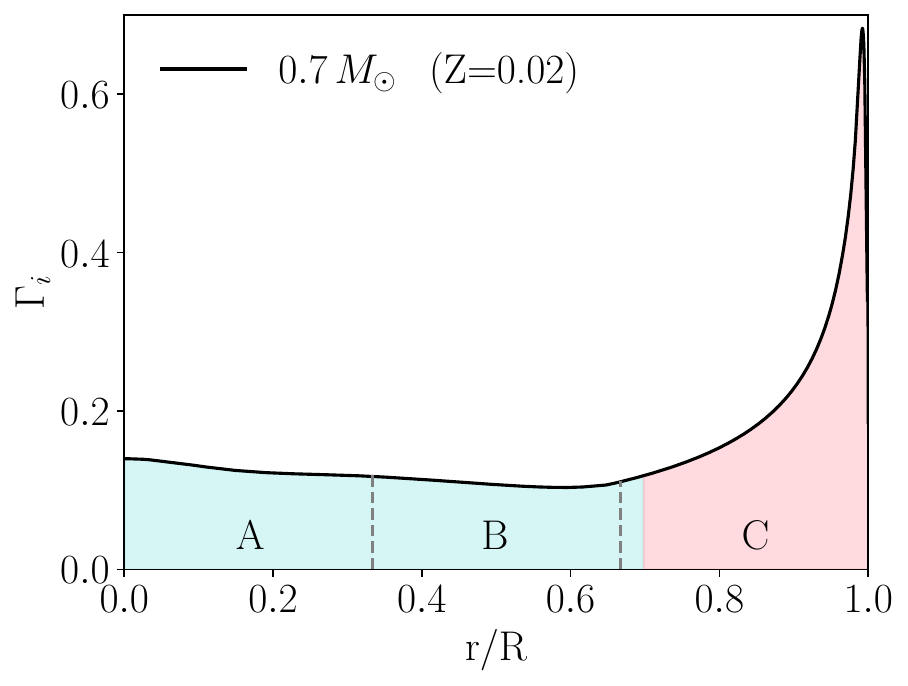}
        \includegraphics[width=6cm]{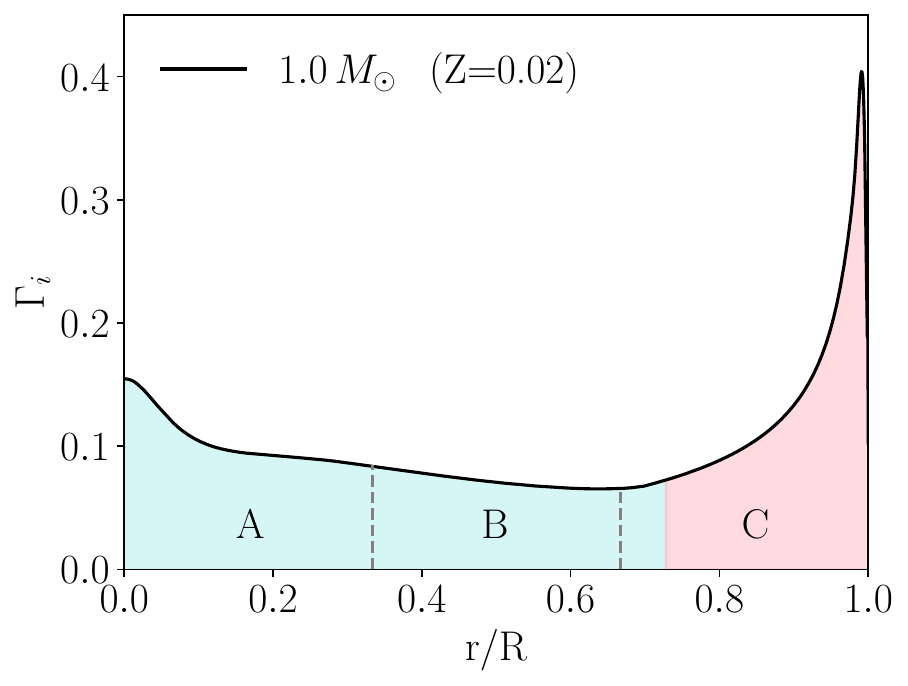}
        \includegraphics[width=6cm]{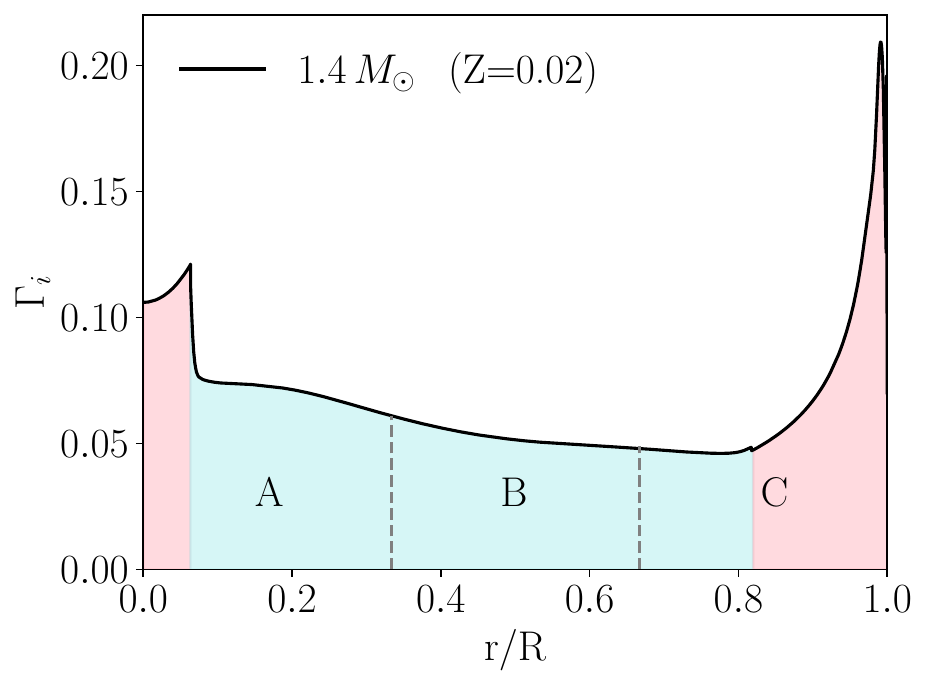}

        \includegraphics[width=6cm]{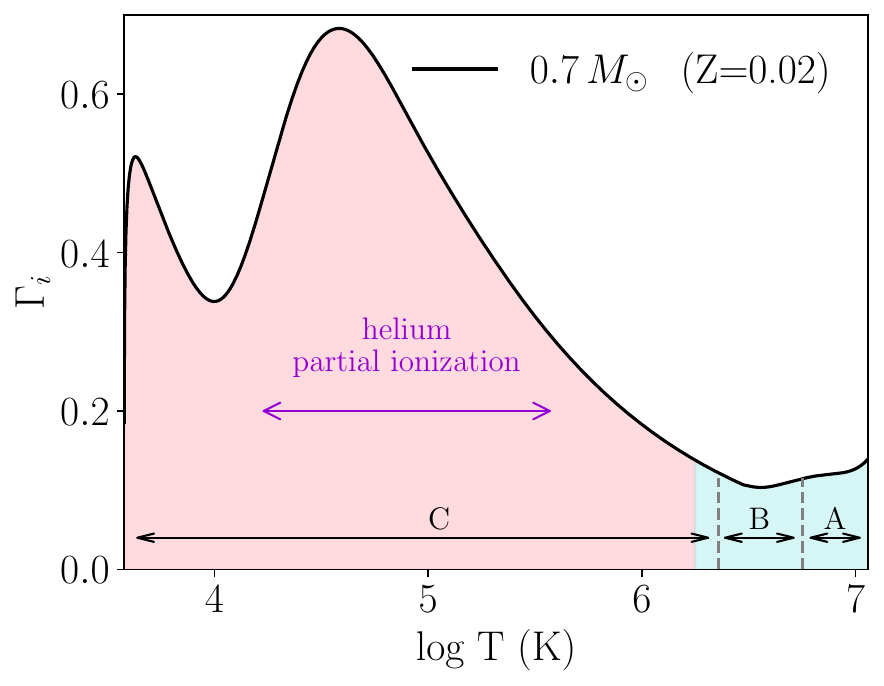}
        \includegraphics[width=6cm]{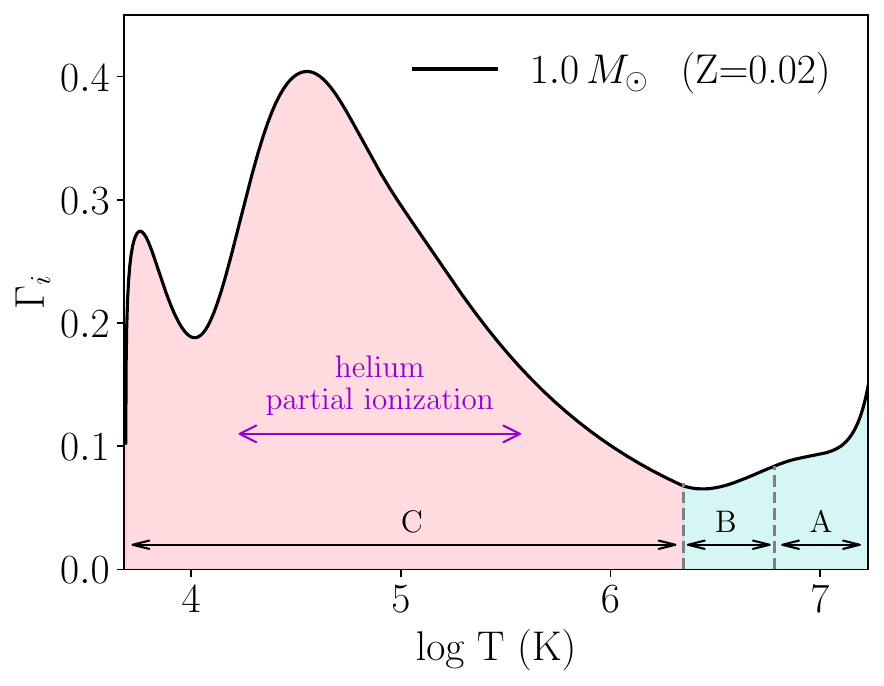}
        \includegraphics[width=6cm]{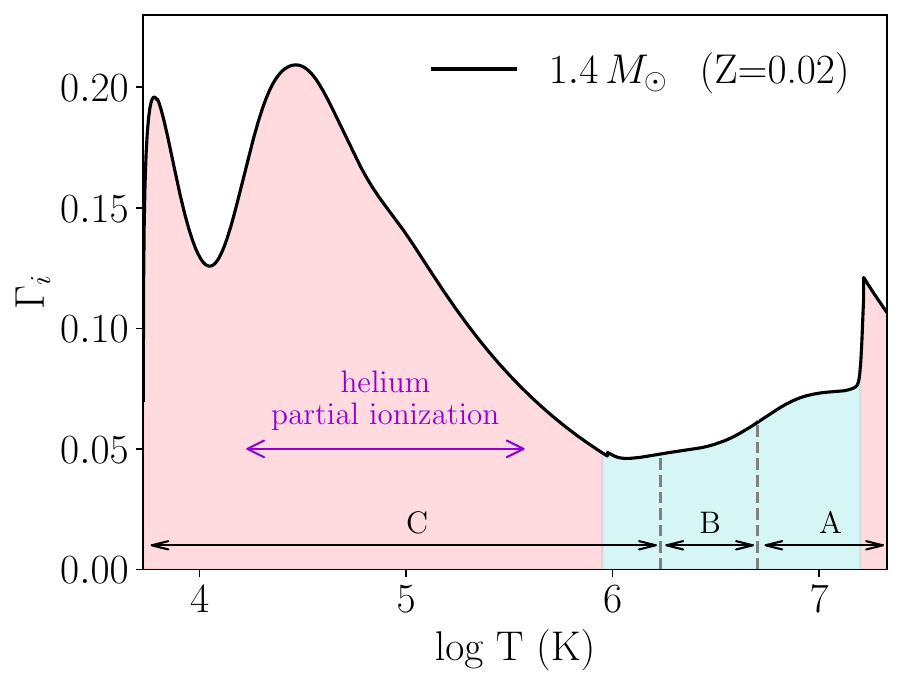}
        \caption{Ionic plasma coupling parameter, $\Gamma_i$, illustrated for three models with varying masses at $Z=0.02$. The upper panel shows $\Gamma_i$ as a function of the fractional radius, while the lower panel shows $\Gamma_i$ as a function of temperature ($\log T$). Regions with higher values of $\Gamma_i$ indicate a greater influence of electrostatic effects within the stellar interior. Across these three models, the significance of Coulomb effects is more pronounced in the outer stellar layers. The helium ionization zones are indicated with a purple double arrow, where the plasma coupling parameter reaches its maximum value. Additionally, three regions of equal length in radius are shown: Region A corresponds to the innermost third of the model, Region B to the central third, and Region C to the outermost third of the stellar interior. In all the plots, light blue regions correspond to radiative zones, and light red regions to convective regions. Finally, all three models in this figure are at IAMS, meaning they are in the middle of their evolutionary path on the MS.}
        \label{fig3}
\end{figure*}

We based our grid of theoretical models on a set of MS stars taken from the catalog of dwarf stars with asteroseismic rotation rates \citep{2021NatAs...5..707H}. This dataset contains a total of 94 stars. In the catalog, these 94 stars are classified into three groups by ``type.'' \citet{2021NatAs...5..707H} identified three different types: ``subgiants'' (SG) with only 4 stars, ``hot stars'' (H) with 24 stars, and ``main sequence'' (MS) with 66 stars. The stars we selected, represented in red in Figure \ref{fig1}, are the 66 stars from the catalog clearly classified as MS. These are 66 low-mass stars with oscillation frequencies  detected at  high signal-to-noise ratios.
Thus, these stars, which serve as a basis for our theoretical models, have $T_{\text{eff}} < 6250$ K and  $ \text{log} \, g > 4$. Specifically, we considered all stars in the catalog of \citet{2021NatAs...5..707H} except those classified as H and SG.

The idea is to cover the $\log g - T_{\text{eff}}$ range of the subsample of 66 MS stars. We did not intend to model the stars to exactly match the $\log g - T_{\text{eff}}$ values of the observed stars. Our goal is to study the electrostatic properties of stellar interiors using a set of models for which we know there are observed stars with similar properties in terms of mass, age, and metallicity. We proceeded as follows.
First, we computed a grid from 0.7 to 1.4 $M_\odot$ (with a step of 0.1) as this is the stellar mass range of the subsample of observed stars. The age of the models is based on the average age of the observed stars for each mass category. Here, ``mass category'' refers to all the stars in the selected subsample with mass values rounded to the closest model mass values (0.7, 0.8, 0.9, 1.0, 1.1, 1.2, 1.3, or 1.4 $M_\odot$). For each stellar mass, we computed models with the following metallicities: Z=0.005, 0.01, 0.02, 0.03, and 0.04. By doing this, we obtained a set of 40 models.
We then allowed some of the models from this initial set to vary in age to fill the gaps in the $\log g - T_{\text{eff}}$ values of the observed stars. This procedure resulted in 37 models. Our final set of models comprises 77 models. 

Figure \ref{fig1} shows the distribution of our models in the $\log g - T_{\text{eff}}$ diagram. In this Kiel diagram, the observational values for the set of 66 Kepler stars are shown in red. The models that extend to lower and higher effective temperatures and fall outside the $\log g - T_{\text{eff}}$ values of the observed stars are, in general, models with either higher masses and low metallicities or lower masses and high metallicities. As these models are interesting for the study of electrostatic properties, we decided to keep them in the study.

In the top panel of Figure \ref{fig1}, the theoretical data are colored according to the ages of the models, whereas in the bottom panel, the same model data are colored according to metallicity. Table \ref{tab1} summarizes some basic statistical details to help the reader better understand the two datasets. Specifically, we compare the average values and their standard deviation (SD) for stellar parameters such as mass, age, metallicity, effective temperature, and surface gravity.

Concerning the input physics, our choices are as follows. Theoretical models were computed with the stellar evolution code Modules for Experiments in Stellar Astrophysics  \citep[MESA v15140;][]{Paxton2011, Paxton2013, Paxton2015, Paxton2018, Paxton2019, 2023ApJS..265...15J}. The code was compiled using \href{https://doi.org/10.5281/zenodo.4587206}{MESA SDK version 20.12.1}. All models were evolved with the metallicity mixture from \citet{2009ARA&A..47..481A}. The opacities used are from OPAL tables at high temperatures \citep{Iglesias1993, 1996ApJ...464..943I} complemented at low temperatures with opacities from \citet{2005ApJ...623..585F}. MESA relies on an equation of state that is a blend of several equations of state \citep{Saumon1995, Timmes2000, Rogers2002, Irwin2004, Potekhin2010, Jermyn2021}. The nuclear reaction rates were obtained from JINA REACLIB \citep{Cyburt2010} with included screening effects using the prescription of \citet{Chugunov2007}. Additionally, all models include atomic diffusion according to \citet{1994ApJ...421..828T}, as previous studies describe it as an important element-transport process for low-mass stars \citep[e.g.,][]{2018MNRAS.477.5052N, 2022A&A...666A..43M}. Although radiative accelerations have also been shown to be important at low metallicities and for higher masses within the range studied here \citep[e.g.,][]{2018A&A...618A..10D}, we did not include them in this study. This is because our aim in this initial study of the electrostatic properties of stellar interiors is to keep our models simple and maintain consistent input physics across all models. However, in subsequent studies, it will be very interesting to investigate the impact that radiative accelerations can have on the electrostatic properties of the theoretical models. Convection is treated according to the mixing-length theory from \citet{1958ZA.....46..108B}, without overshoot, and using a mixing-length parameter value of $\alpha_{\text{MLT}} = 1.8,$ common
to all models. Finally, the outermost layers on the models are described by a Grey-Eddington atmospheric structure.
A set of files that allows the reader to reproduce all the models in this work, and consequently all the findings in this study, is openly available through Zenodo with the DOI: 10.5281/zenodo.7749260.

\section{Electrostatic properties of stellar interiors}\label{sec:3}

In this section, we study electrostatic effects in the interiors of the theoretical models represented in Figure \ref{fig1}. We also investigate the variations of these effects with mass, age, and metallicity.

\subsection{The Debye length and the total energy density of the electrostatic interactions}\label{subsec:3.1}

\begin{figure*}
        \centering
        \includegraphics[width=13cm]{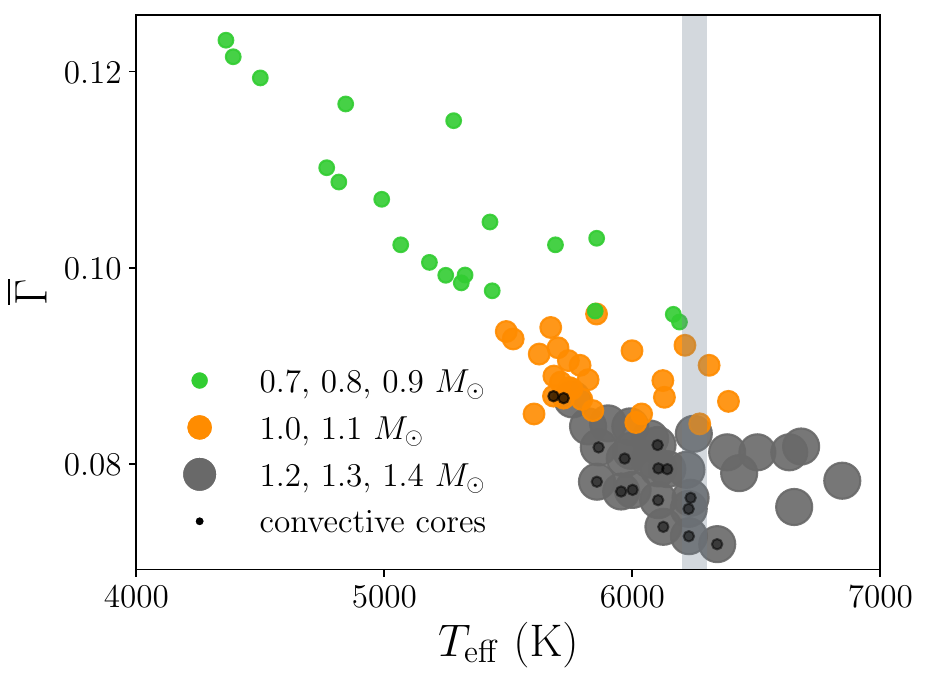}

        \includegraphics[width=5.7cm]{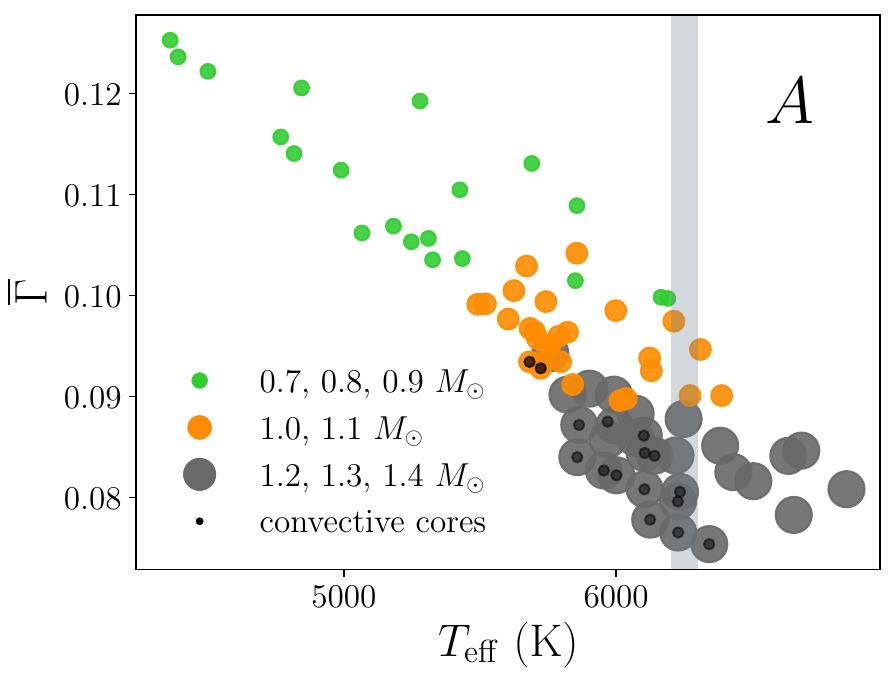}
        \includegraphics[width=5.7cm]{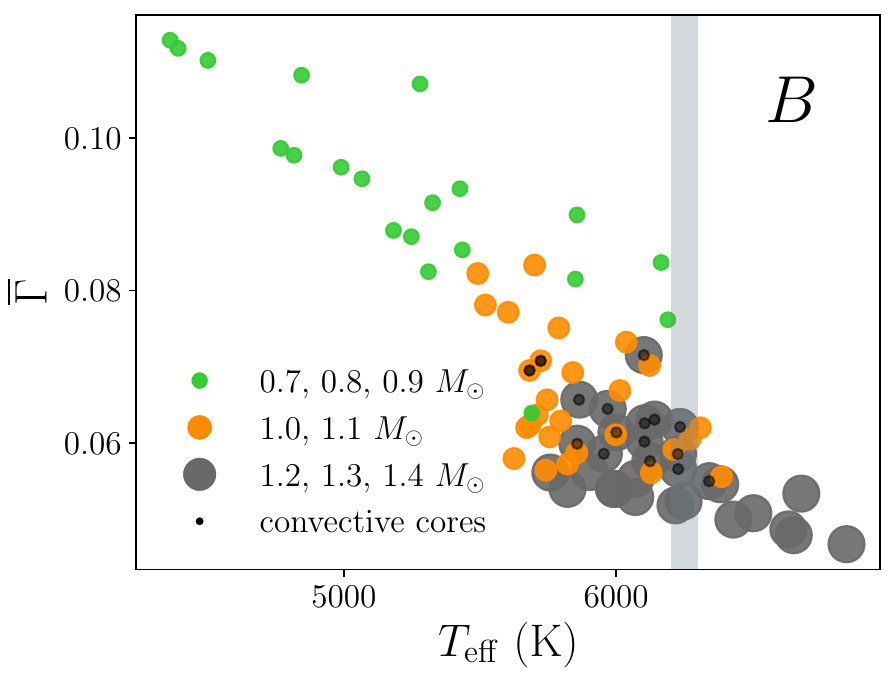}
        \includegraphics[width=5.7cm]{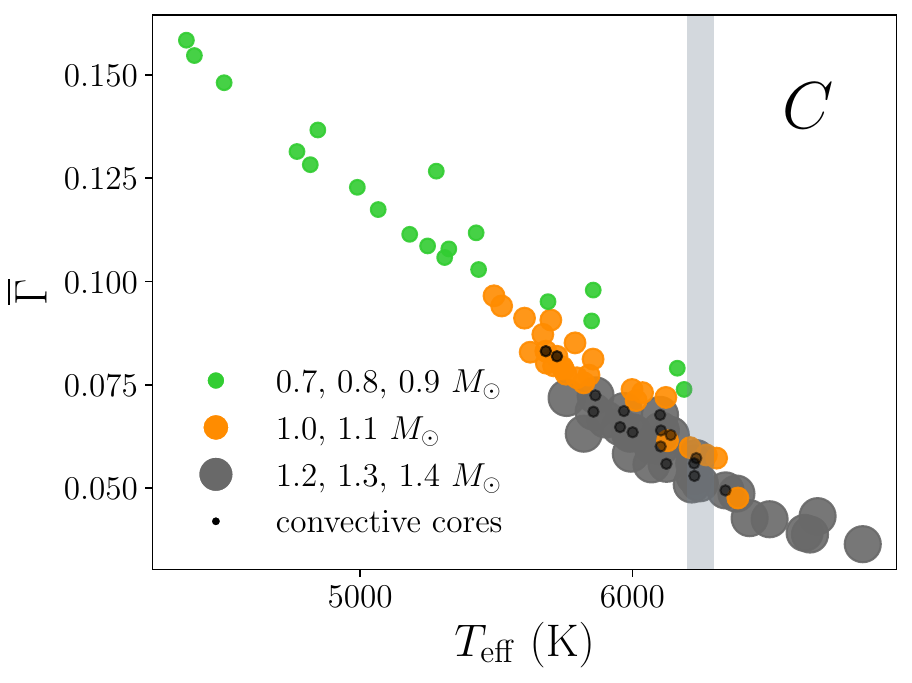}
        \caption{
                 Value of $\overline{\Gamma}$ as a function of the effective temperature for all the stellar models. Upper panel: Value of $\overline{\Gamma}$ computed throughout the entire stellar interior. The green color highlights the less massive models
                (0.7, 0.8, and 0.9 $M_\odot$), whereas the gray color highlights the more massive models (1.2, 1.3, and 1.4 $M_\odot $). The orange color corresponds to models with 1.0 or 1.1 $M_\odot$. The vertical dashed line indicates the approximate location of the Kraft break.
                Lower panel: Value of $\overline{\Gamma}$  computed for three different regions (A, B, and C) within the stellar interiors. Here, we show the computed values as a function of the effective temperature for all stellar models. Regions A, B, and C are described in Figure \ref{fig3} . The green, gray, and orange colors have the same meaning as in the upper panel. The small black circles in all the plots indicate the presence of a model with a convective core.}
        \label{fig4}
\end{figure*}

\begin{figure*}
        \centering
        \includegraphics[width=6cm, height=5cm]{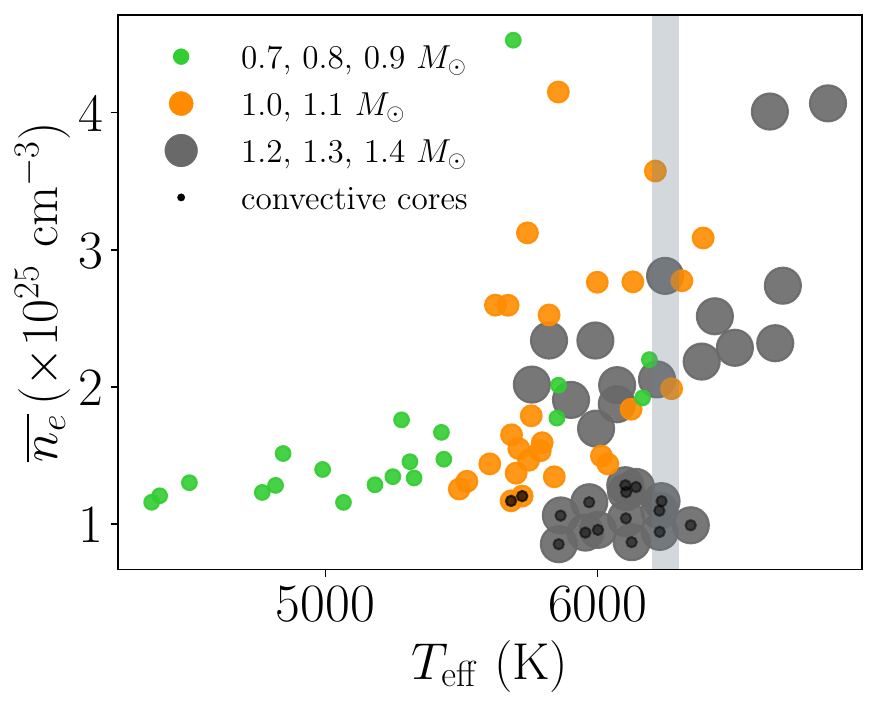}
        \includegraphics[width=6cm, height=5cm]{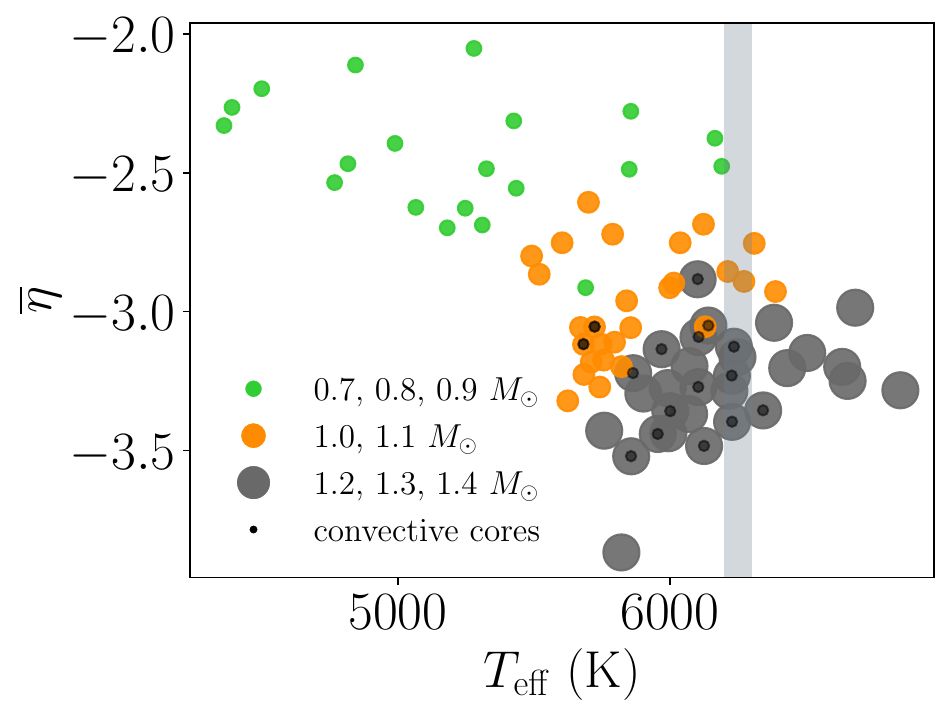}
        \includegraphics[width=6cm, height=5cm]{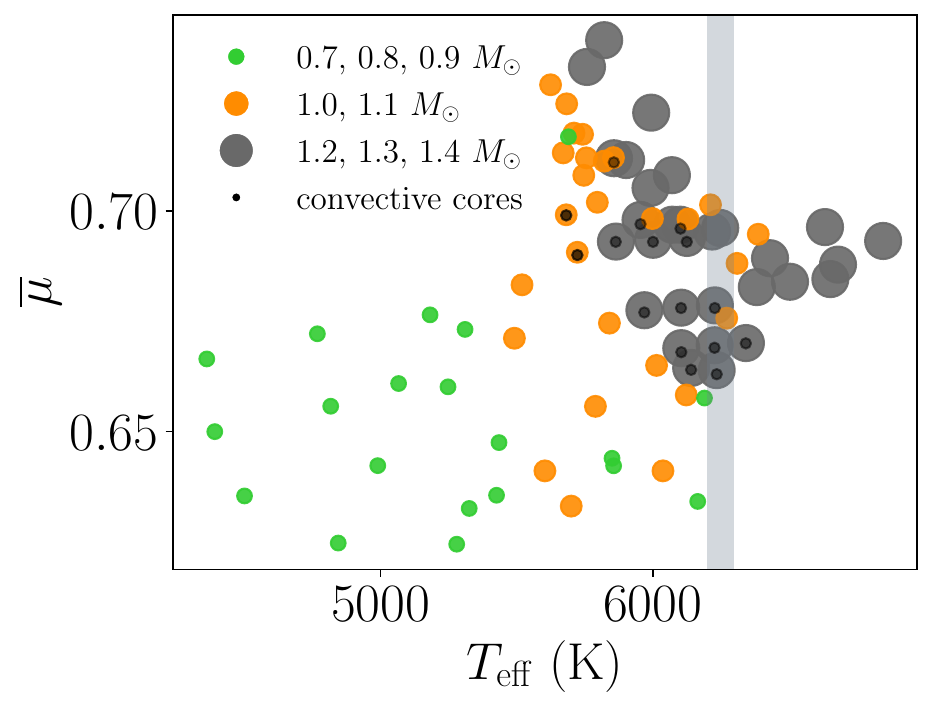}
        \caption{Values of $\overline{n_e}$, $\overline{\eta}$, and $\overline{\mu}$ as a function of the effective temperature for all the stellar models. The green,
gray, and orange colors have the same meaning as in Fig. 4, as do the vertical dashed line  and the small black circles.}
        \label{fig5}
\end{figure*}

In Section \ref{sec:1}, we highlight quasi-neutrality as a key concept defining plasma. Quasi-neutrality is a state that the plasma actively seeks to attain by continuously readjusting the local distribution of charged particles in response to perturbations in charge. These effects are commonly known as plasma screening effects, and were initially developed in the pioneering work of \citet{Debye_1923}. Thus, a crucial property of plasma is that the charged particles arrange themselves in such a way as to shield any electrostatic fields within a certain distance. This distance, usually represented by $\lambda_D,$ is called the Debye length. The Debye length is a measure of the shielding distance.
Let us consider the electric charge density, that is, the net density of electric charges at a given location, $\rho_e$, defined as the difference between the ionic electric charge density and the electronic charge density \citep[e.g.,][]{2004cgps.book.....W}:
\begin{equation}
        \rho_e = \sum_i n_i Z_i e - n_e e .
        \label{eq1}
\end{equation}
Here, the index $i$ represents the different atomic species that constitute the stellar plasma and $e$ is the elementary charge. Boltzmann's distributions developed to first order allows us to write the ionic concentration as
\begin{equation}
        n_i(r) = n_{0i} \left( 1 - \frac{Z_i e \Phi(r)}{kT} \right) \, ,
        \label{eq2}
\end{equation}
as well the concentration of electrons as
\begin{equation}
        n_e(r) = n_{0e} \left( 1 + \frac{e \Phi(r)}{kT} \right) \, .
        \label{eq3}
\end{equation}
The values $n_{0i}$ and $n_{0e}$ represent, respectively, the concentrations of ions and electrons when the stellar matter is unperturbed. Because unperturbed matter is neutral, we have $n_{0i}Z_i = n_{0e}$ which represents the well-known neutrality condition. Finally, $\Phi(r)$ is the electric potential, which is given by the Poisson equation
\begin{equation}
        \nabla^2 \Phi(r) = - 4 \pi \rho_e .
        \label{eq4}
\end{equation}
The Poisson equation relates $\Phi(r)$ to the number densities $n_i (r)$ and $n_e (r)$ \citep[e.g.,][]{1998asa..book.....R, 1999stma.book.....M}. Specifically, by substituting the ionic and electronic densities, given respectively by Equations \ref{eq2} and \ref{eq3}, into Equation \ref{eq4}, it is possible to obtain an expression for the electrostatic potential of a charge $Ze$:
\begin{equation}
        \Phi(r) = \frac{Ze}{r} e^{-r/\lambda_D} \simeq \frac{Ze}{r} - \frac{Ze}{\lambda_D} .
        \label{eq5}
\end{equation} 
This is the potential that takes into account the fact that electrons tend to surround an ion of positive charge $Ze$. The quantity $\lambda_D$ is the Debye length and can be written as
\begin{equation}
        \lambda_D = \sqrt{\frac{kT}{4 \pi e^2 (n_e + \sum_i n_i Z_i^2)}} .
        \label{eq6}
\end{equation}
This length scale represents a distance over which the thermal fluctuations of the stellar material can lead to an important separation between negative and positive charged particles \citep[e.g.,][]{1999stma.book.....M}.

We computed the internal profiles of the Debye lengths for several of our models. In Figure \ref{fig2} (left panel), we show how the Debye length varies with mass for Z=0.02 metallicity models at an intermediate age on the main sequence (IAMS), which means that the age of the model is $50 \%$ of the entire MS lifetime. Typically, for the same age and metallicity, the Debye length increases as the mass of the stellar model also increases.
Debye lengths also vary with the age of a star. The central panel of Figure \ref{fig2} shows the variation of $\lambda_D$ along the MS for the less massive and more massive models. As the stars evolve on the MS, the Debye length increases with the mass of the stellar model by a few orders of magnitude.

Screening effects, and in particular the screening potential (Equation \ref{eq5}),  have a significant influence on the microscopic and macroscopic properties of stellar interiors \citep[e.g.,][]{2008PhR...457..217B, 2004ApJ...614..464B}. One important consequence is the reduction of pressure in the stellar medium as a result of screening effects. This occurs because a charge produces a surrounding cloud of radius $\lambda_D$ containing an excess of charges with opposite sign. The system formed by the charge and the cloud is bound and electrically neutral, and has a negative energy as it is necessary to provide energy to separate it. This negative energy leads to a pressure decrease in the stellar interior. The total energy density of electrostatic interactions depends on the Debye length, and can be written as \citep{2009pfer.book.....M}:
\begin{equation}
        u_{ES} = - \frac{1}{2} \frac{e^2}{\lambda_D} \sum_i n_i Z_i^2 \, .
        \label{eq7}
\end{equation}
This energy density is plotted for different stellar masses in the right panel of Figure \ref{fig2} . All models represented in Figure \ref{fig2} were computed at Z=0.02 metallicity and have evolved up to the MS halftime, with the exception of two models in the central panel, represented by dashed lines, which were computed at zero age main sequence (ZAMS).


\begin{table}
        \caption{Ages of the models at Z=0.02 metallicity}              
        \label{tab2}      
        \centering                                      
        \begin{tabular}{c c c}          
                \hline\hline                        
                Mass & Age at ZAMS (Gyr) & Age at TAMS (Gyr) \\    
                \hline                                   
                $0.7 \, M_\odot$ & 0.0770 & 38.3 \\
                $0.8 \, M_\odot$ & 0.0603 & 24.4 \\
                $0.9 \, M_\odot$ & 0.0486 & 16.0 \\
                $1.0 \, M_\odot$ & 0.0381 & 10.6 \\
                $1.1 \, M_\odot$ & 0.0298 & 7.40 \\
                $1.2 \, M_\odot$ & 0.0239 & 5.34 \\
                $1.3 \, M_\odot$ & 0.0195 & 3.96 \\
                $1.4 \, M_\odot$ & 0.0162 & 3.05 \\
                \hline                                             
        \end{tabular}
        \tablefoot{
        This table displays the ages of the models represented in Figure \ref{fig6}. Specifically, the ages of the models at ZAMS and the ages of the models at the chosen TAMS.
        }
\end{table}

\begin{figure}
        \centering
        \includegraphics[width=9cm]{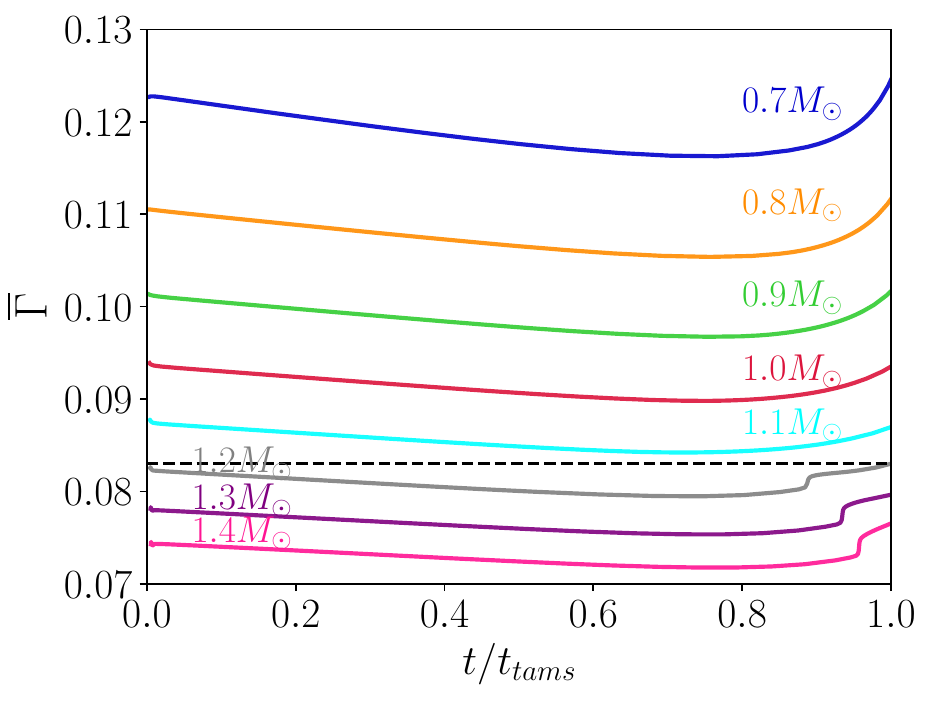}
        \caption{Variation of the global plasma parameter, $\overline{\Gamma}$ , as a function
                of the normalized age for stellar models with different masses. All the
                models represented in this figure were computed with $Z=0.02$.
                Lower-mass stellar models exhibit higher values of the global plasma
                parameter throughout the MS lifetime. The dashed black line
                represents the approximate location of the Kraft break.}
        \label{fig6}
\end{figure}

\subsection{The plasma coupling parameter}\label{subsec:3.2}

The occurrence of the Debye shielding effect is usually a first example of a plasma collective behavior. It is easy to see from Equation \ref{eq6} that, as density increases, the Debye length decreases. Furthermore, as the thermal kinetic energy, $kT$, of the particles increases, the Debye length increases as well. As a consequence, the properties and the particular characteristics of the stellar plasma will depend on two quantities: the electrostatic potential energy, and the thermal energy of the particles that constitute the stellar material (electrons and ions). 

An important ratio that allows us to characterize the electrostatic properties of stellar interiors is the plasma coupling parameter. This parameter, represented here by $\Gamma_i$, is defined as the ratio of Coulomb potential energy to the thermal energy. Considering the ionic case, $\Gamma_i$ can be written as
\begin{equation}
        \Gamma_i = \frac{(\overline{Z_i} e)^2}{a_i k T}\, ,
        \label{eq8}
\end{equation}
where the Wigner-Steitz radius, $a_i$, is given by
\begin{equation}
        a_i = \left( \frac{3 \overline{Z_i}}{4 \pi n_e}\right)^{1/3}
        \label{eq9}
,\end{equation}
and represents the ion sphere radius, or in another words, the mean inter-ion distance \citep[e.g.,][]{1986ApJS...61..177P, 2009pfer.book.....M}. Here, $e$ is the elementary charge, $k$ the Boltzmann constant, $T$ the local temperature, $\overline{Z_i}$ the mean ionic charge, and $n_e$ the electron density.
The higher the value of the plasma coupling parameter, the more important the Coulomb interactions between ions. When Coulomb and thermal energies are balanced, the plasma coupling parameter takes values of around unity. For cases where electrostatic interactions dominate over thermal energies, the plasma coupling parameter becomes greater than one. 
It is well known that, for low-mass stars, Coulomb effects should be considered in the equation of state \citep[e.g.,][]{2021LRSP...18....2C}. Moreover, a recent study \citep{2021MNRAS.507.5747B} showed the particular significance of electrostatic interactions in the ionization zones of the most abundant elements for low-mass stellar models.

Figure \ref{fig3} displays the ionic plasma coupling parameter plotted for three of our theoretical stellar models ($0.7 \, , 1.0 \,$ and $1.4 \, M_\odot$), with all three computed at $Z=0.02$. We note the clear increase in the importance of electrostatic effects as the mass of the stellar model decreases, with $\Gamma_i$ peaking near the surface for all models, a known behavior of the plasma parameter \citep[e.g.,][]{2021LRSP...18....2C}. The lower panel in Figure \ref{fig3} shows that the maximum values of the plasma coupling parameter are reached within the temperature interval where the two helium ionizations occur. This figure highlights the fact that specific aspects of the stellar structure, particularly the underlying microphysics of the outer convective zones, are crucial for understanding the electrostatic properties of low-mass stellar interiors. In Figure \ref{fig3}, convective zones are represented in light red, while radiative zones are shown in light blue.

\begin{figure*}
        \centering
        \includegraphics[width=8cm]{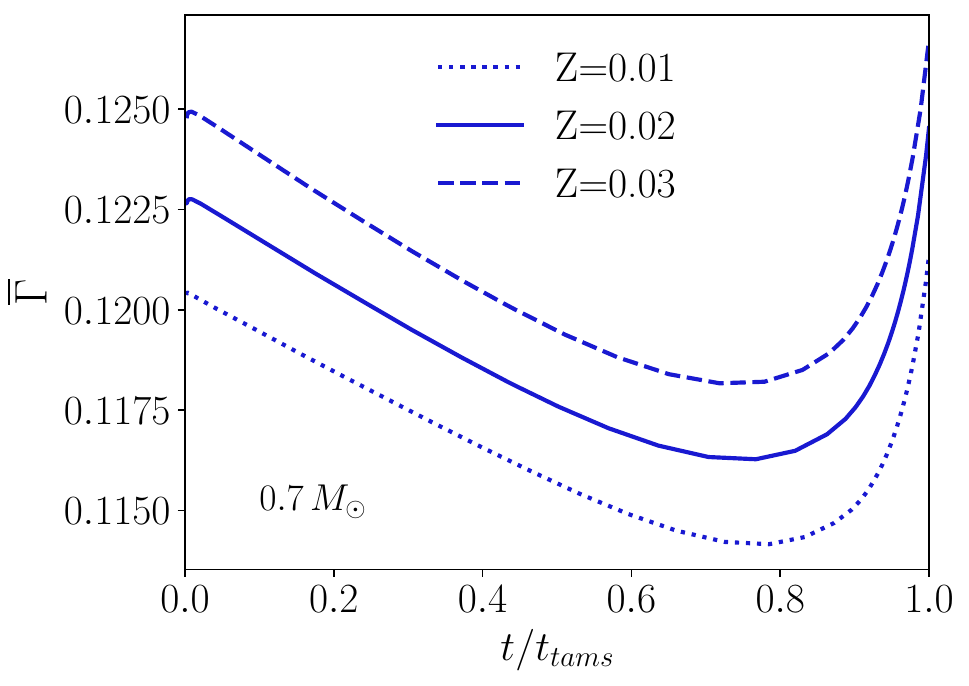}
        \includegraphics[width=8cm]{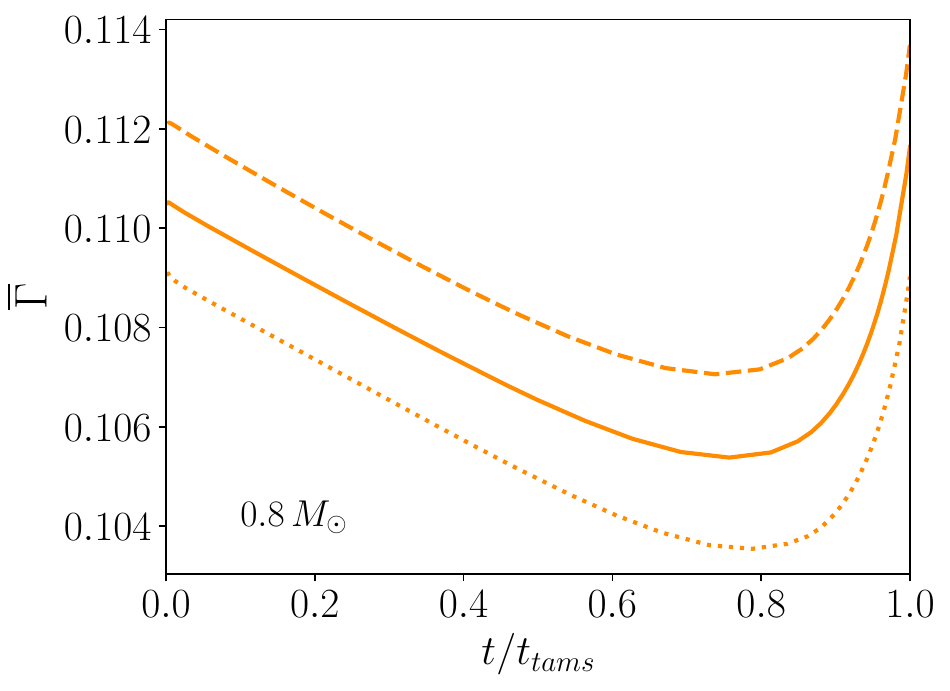}

        \includegraphics[width=8cm]{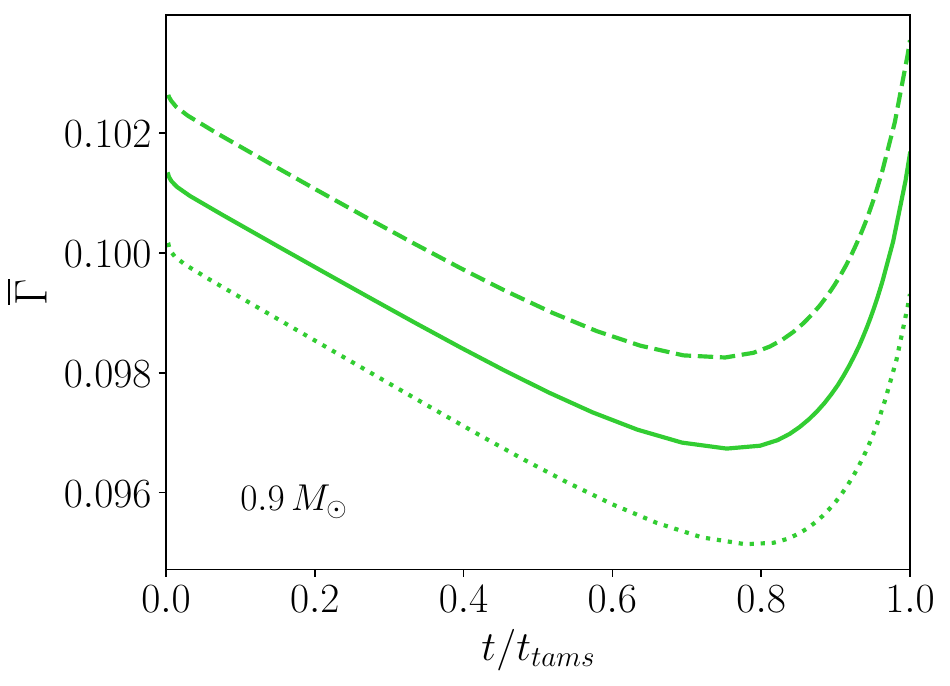}
        \includegraphics[width=8cm]{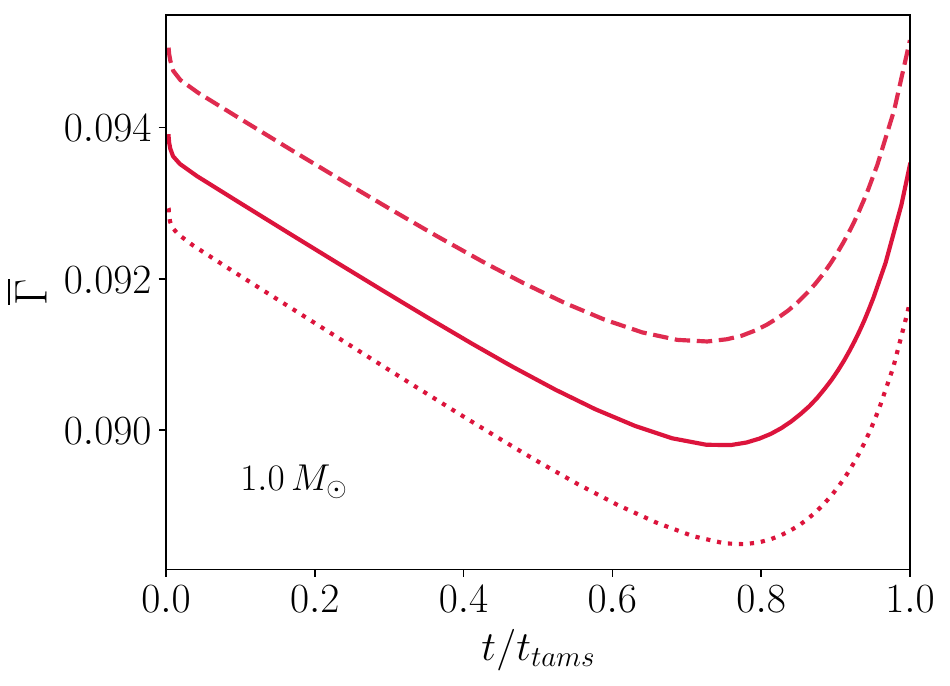}

        \includegraphics[width=8cm]{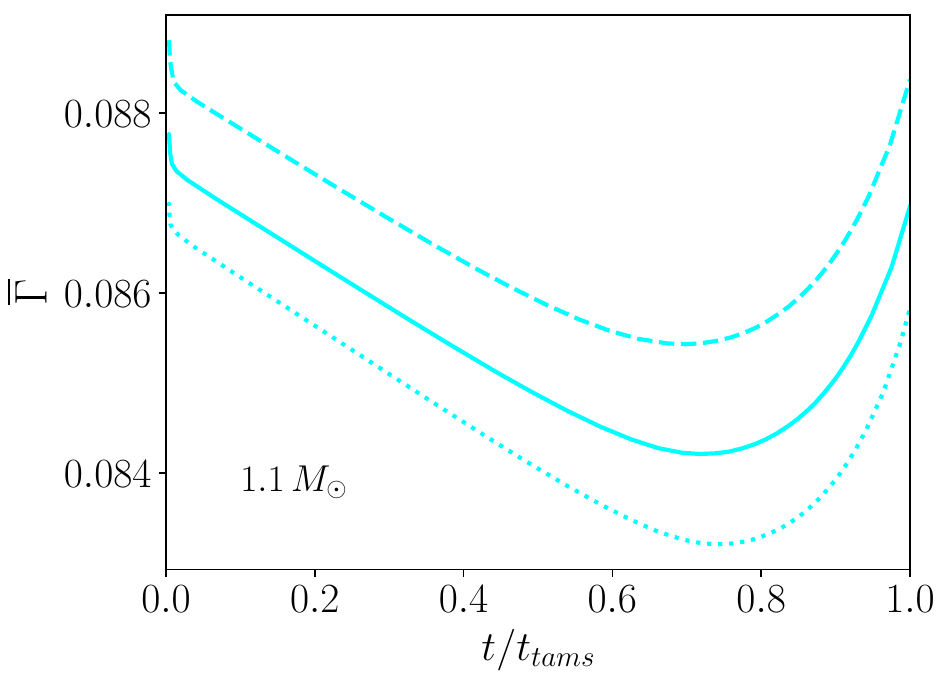}
        \includegraphics[width=8cm]{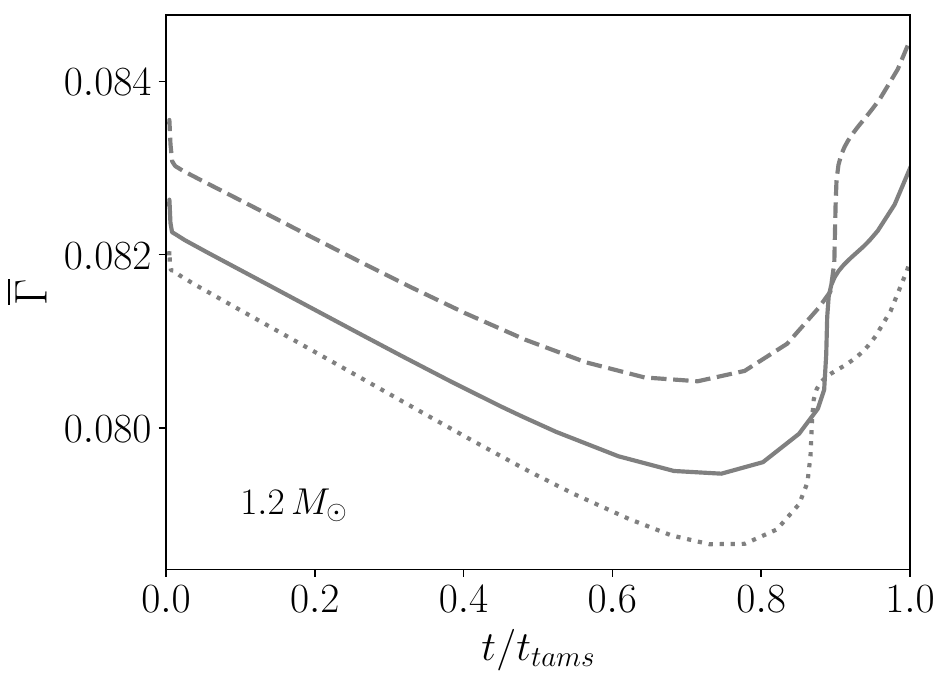}

        \includegraphics[width=8cm]{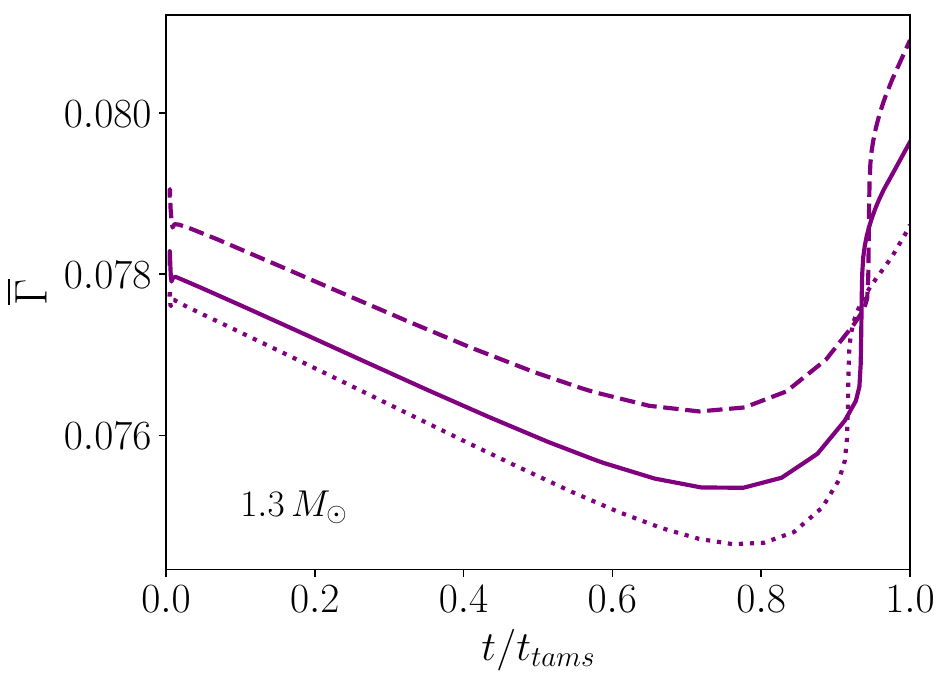}
        \includegraphics[width=8cm]{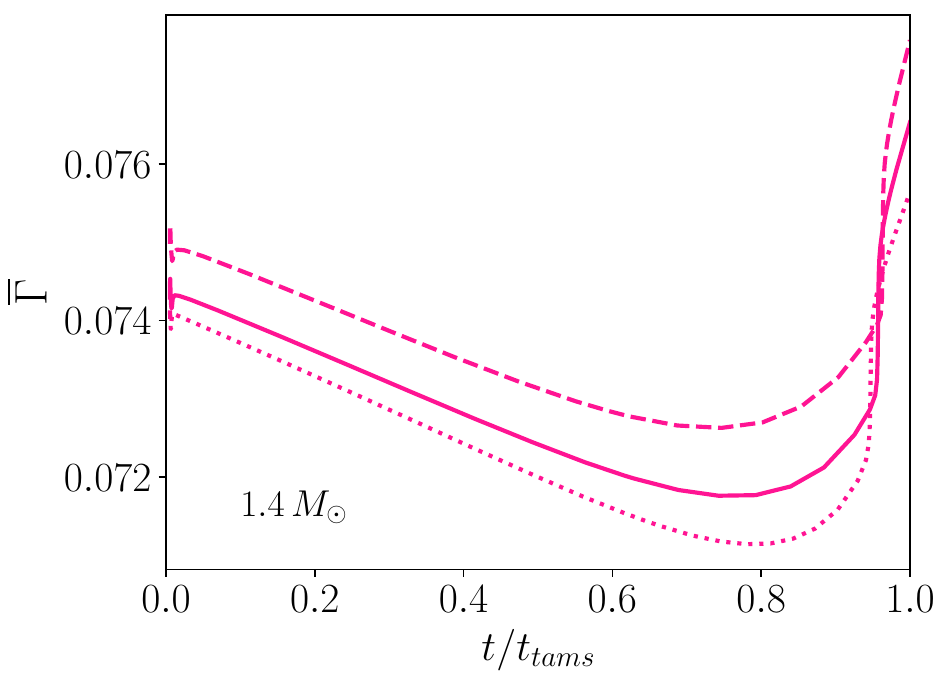}
        \caption{Global plasma parameter, $\overline{\Gamma}$, as a function of age for different metallicities. The solid lines show the same models as those represented in Figure \ref{fig6}, i.e., models computed at Z=0.02 metallicity. Dashed lines represent high-metallicity models (Z=0.03), whereas dotted lines represent low-metallicity models (Z=0.01).}
        \label{fig7}
\end{figure*}

\subsection{Global stellar plasma parameter}\label{subsec:3.3}

The plasma coupling parameter is a local variable that characterizes the balance between electrostatic and thermal energies at each location in the stellar interior. Nevertheless, by considering a mean value of the parameter $\Gamma_i$, defined by the relation
\begin{equation}
        \overline{\Gamma} = \frac{1}{M} \int_{0}^{M} \Gamma_i \, dM(r) \, ,
        \label{eq10}
\end{equation}
where
\begin{equation}
        dM(r)= 4\pi r^2 \rho \,  dr \, ,
        \label{eq11}
\end{equation}
it is possible to define a new variable $\overline{\Gamma}$, which carries a global character. This new global variable can be used to characterize a stellar model from the viewpoint of the importance of Coulomb interactions. The greater the value of $\overline{\Gamma}$, the greater the significance of the Coulomb interactions for the star as a whole. We call this new variable, $\overline{\Gamma}$, the global mean plasma parameter, or simply, the global plasma parameter. From Figure \ref{fig3}, we are able to see that we can use this global variable to study the importance of electrostatic effects in specific regions within the star. 

For example, the stellar interior can be divided in three parts: a deeper region that includes the stellar core (the inner $33.33\%$ of the stellar radius), a central radiative region (the central $33.33\%$ of the stellar radius), and an upper region that contains the bulk of the convective zone of the stellar models (the external $33.33\%$ of the stellar radius). All our models have an outer convective zone and a radiative interior, with 16 models including a convective core. The outer convective zones of the models range in depth from 3\% to 39\% of the stellar outer layer, while the convective cores vary in thickness from 4.5\% to 7.5\% of the innermost layer. A natural choice for dividing the stellar structure would be to consider the radiative and convective regions. However, because more massive stars have a very thin convective zone, we considered the possibility that such a division could introduce a bias in the comparison of mean plasma parameter values.
We also note that Region C (see Figure \ref{fig3}), the outermost region, includes the convective upper zone of all models and could thus be considered an approximation of a region containing the convective zone. Additionally, we were interested in examining a region that includes the core of the model (not just the convective core), where energy is generated through nuclear reactions. Given these considerations, we ultimately decided to divide the structure into three regions of equal length for this initial study of the electrostatic properties of stellar interiors.

Figure \ref{fig4} shows the dependence of $\overline{\Gamma}$ on the effective temperature for all our models described in Section \ref{sec:2}. 
We observe a clear scaling of $\overline{\Gamma}$ that results in three almost distinct clusters of stars according to their $\overline{\Gamma}$ value. As expected, cooler stars have higher values of $\overline{\Gamma}$, whereas hotter stars exhibit lower values of $\overline{\Gamma}$. The cluster of stars with intermediate $T_{\text{eff}}$ values correspond to stars with the middle $\overline{\Gamma}$ values. This behavior of the $\overline{\Gamma}$ -- $T_{\text{eff}}$  relation mirrors the observational behavior of the $P_{\text{rot}}$ -- $T_{\text{eff}}$ dependence, in the sense that cooler stars have higher values of the rotation period, whereas hotter stars exhibit lower values of the rotation period, with the faster-rotating hot stars usually being those above the Kraft break. The Kraft break \citep{1967ApJ...150..551K} is an observational feature characterized by a steep variation in stellar rotational velocities occurring over a small mass (or temperature) range. The approximate location of this transition is around 6200-6300 K \citep[e.g.,][]{1997ApJ...480..303K, 2013ApJ...776...67V, 2016Natur.529..181V, 2020A&A...636A..76S, 2023ApJ...952..131M, 2023MNRAS.526.4787R}, and it is represented in Figure \ref{fig4} by a vertical gray bar.

As we know, a higher value of $\overline{\Gamma}$ indicates that the electrostatic interactions are more important, and thus collective effects are more significant. These are theoretical models that represent stars located below the Kraft break, which are stars that experience spin down along the MS according to the experimental data. Instead, hotter stars can be associated with lower values of $\overline{\Gamma}$.  The fact that $\overline{\Gamma}$ has a low value means that the electrostatic interactions are less important, making collective effects also less significant. In this case, the theoretical models represent stars that are above the Kraft break, and thus do not experience spin down on the MS. These stars do not lose angular momentum and retain the rapid rotation rates typical of the ZAMS. Therefore, we can say that the properties of the global plasma parameter correlate with the observed rotational dependence on mass and effective temperature. 

The bottom panel of the three plots in Figure \ref{fig4} also shows $\overline{\Gamma}$ as a function of effective temperature for all the stellar models used in this study, but in this case, the global plasma parameter is computed for three specific regions within the star as illustrated in Figure \ref{fig3}. These plots are interesting as they show that the different regions of the stellar interior appear to contribute differently to the total value of the global plasma parameter: while the upper layers of the stellar models contribute strongly to the linear scaling obtained in Figure \ref{fig4} , the inner regions are responsible for a dispersion around the linear scaling. 

The global plasma parameter, which serves as an indicator of the average Coulomb coupling strength within the stellar interior, emerges as a characteristic of the interior that could result in observable effects. From a physical standpoint, the connection between the global plasma parameter and observable rotational patterns appears logical. This is because the plasma coupling parameter quantifies the extent to which many-body interactions influence the dynamics of the plasma. Consequently, it acts as a bridge between the microphysics of the stellar interior and the observable characteristics of the star.

\begin{figure*}
        \centering
        \includegraphics[width=6.0cm]{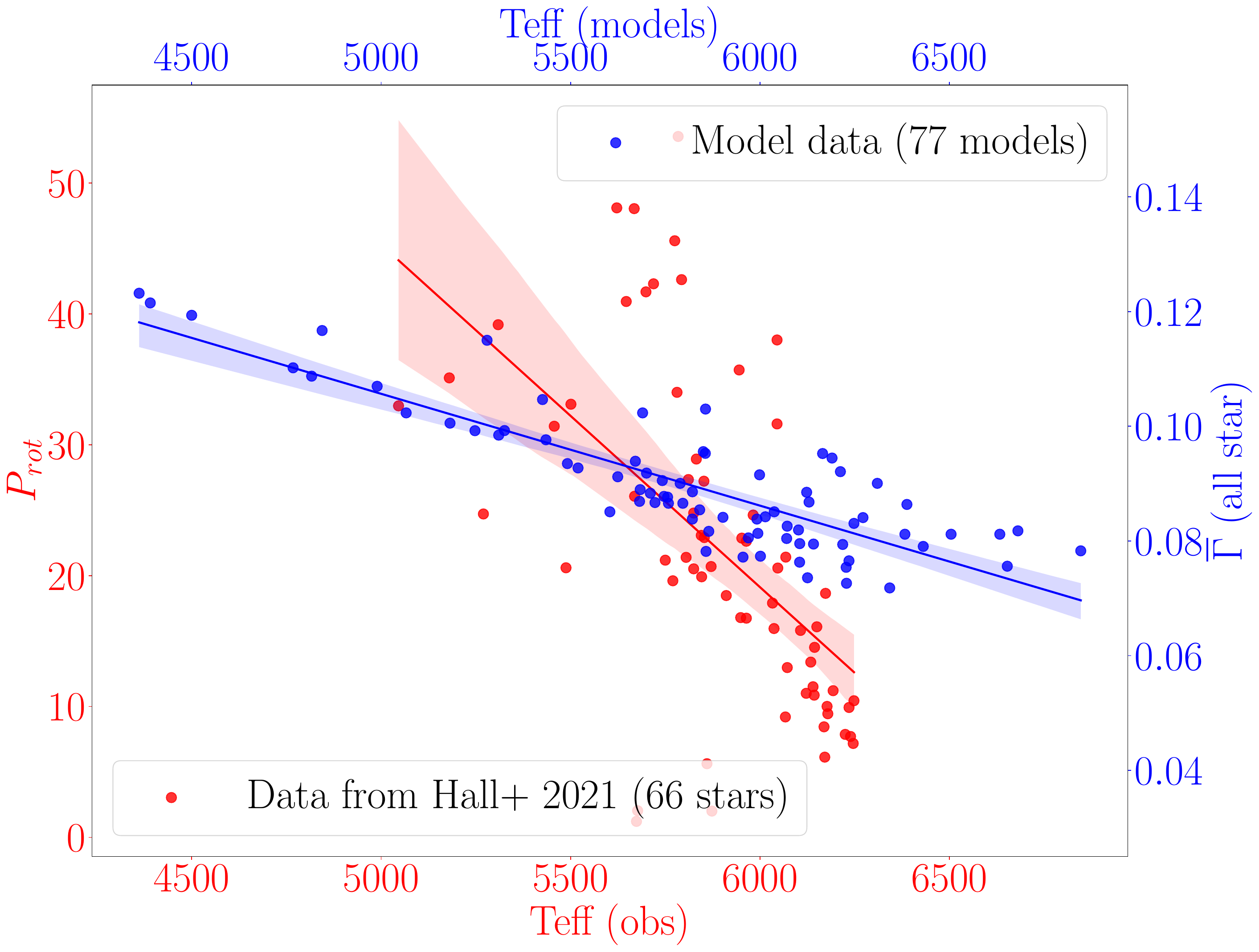}
        \includegraphics[width=6.0cm]{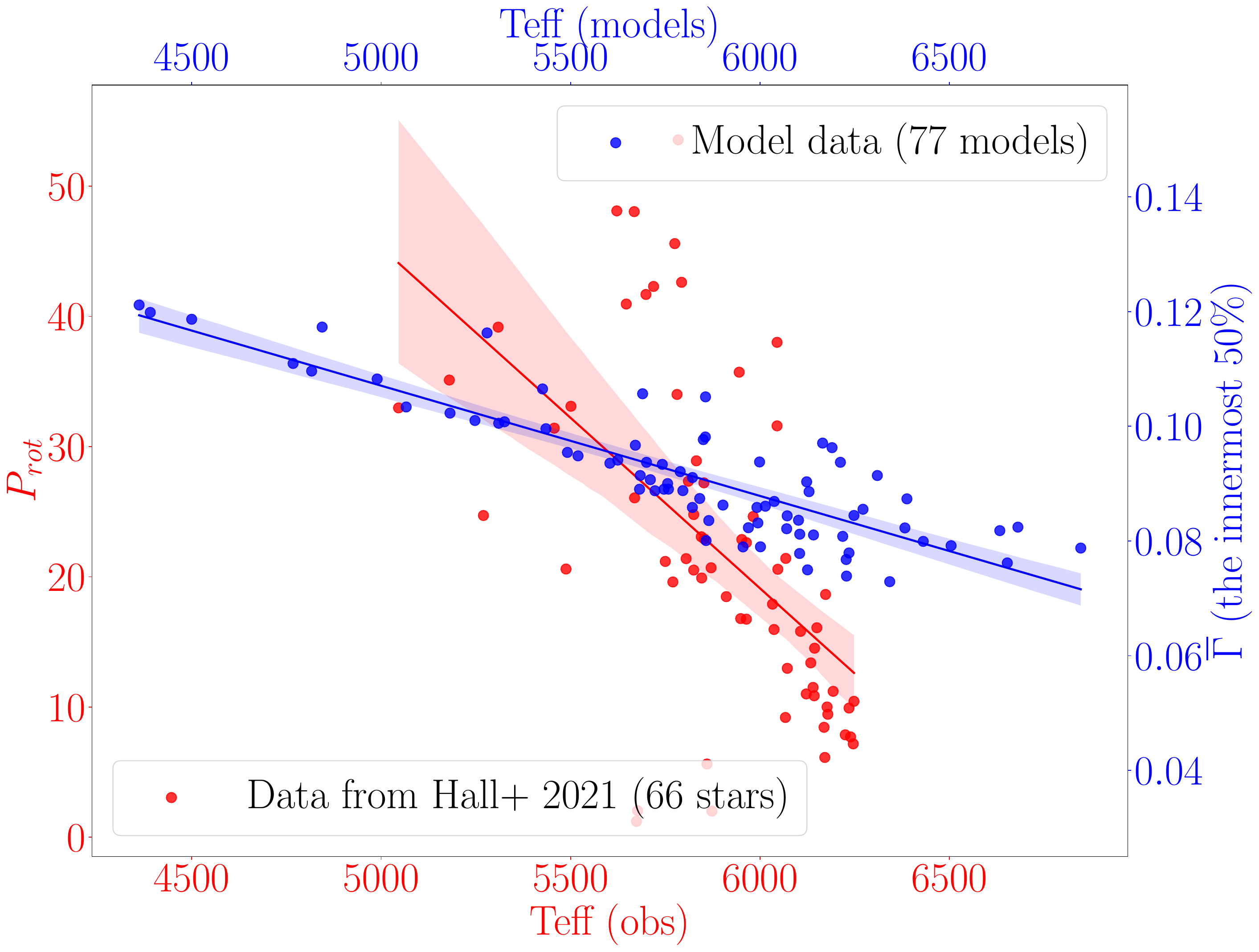}
        \includegraphics[width=6.0cm]{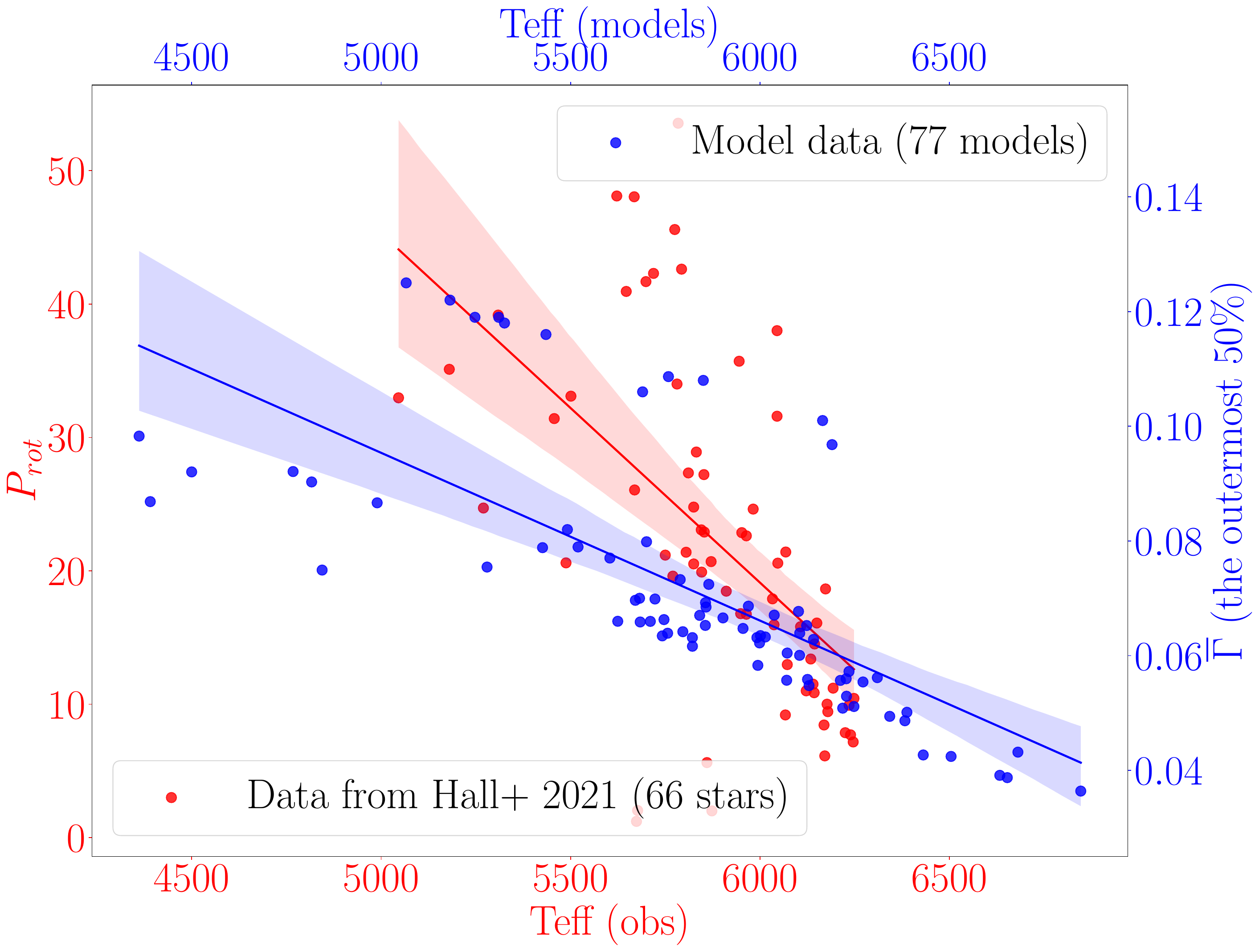}

        \includegraphics[width=6.0cm]{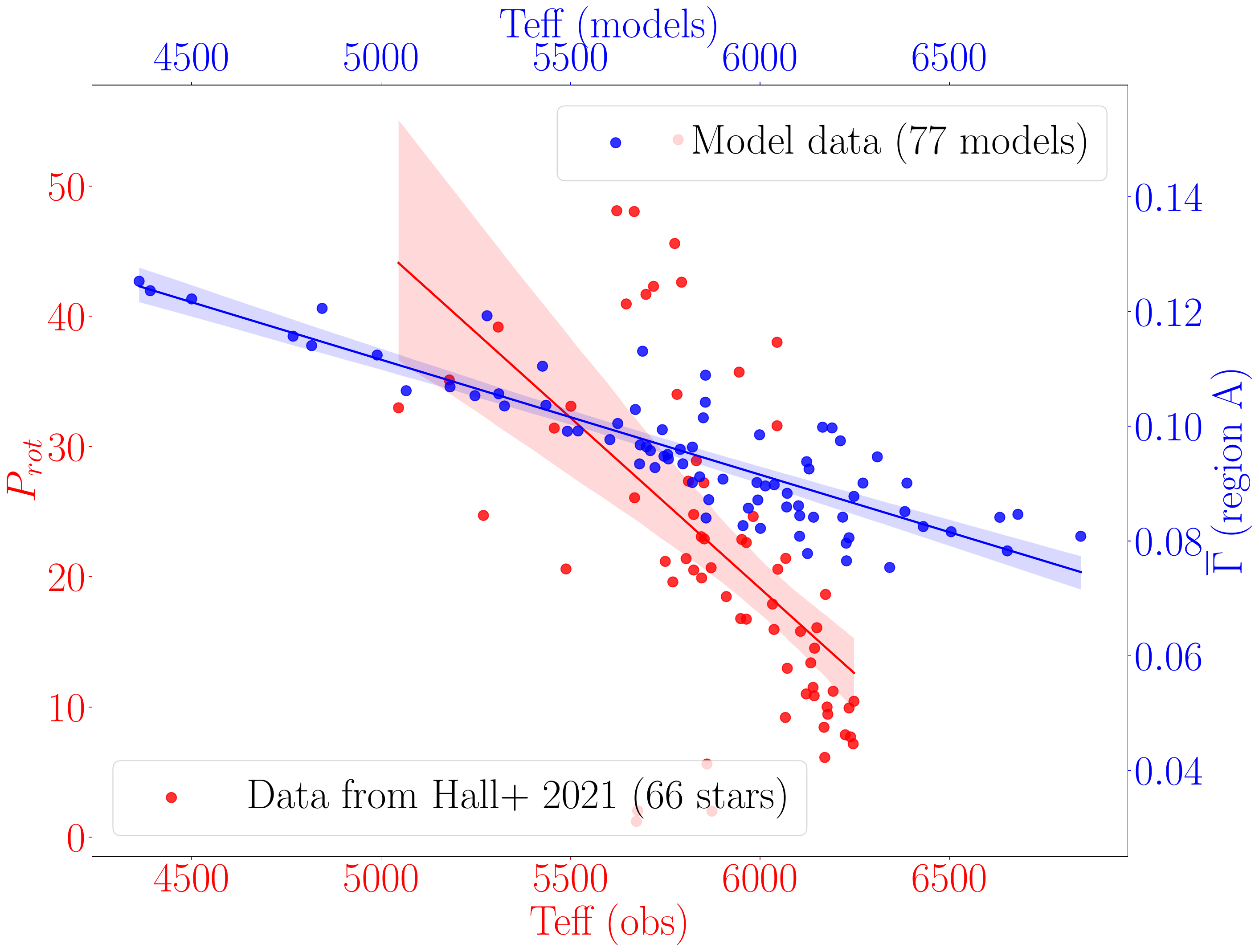}
        \includegraphics[width=6.0cm]{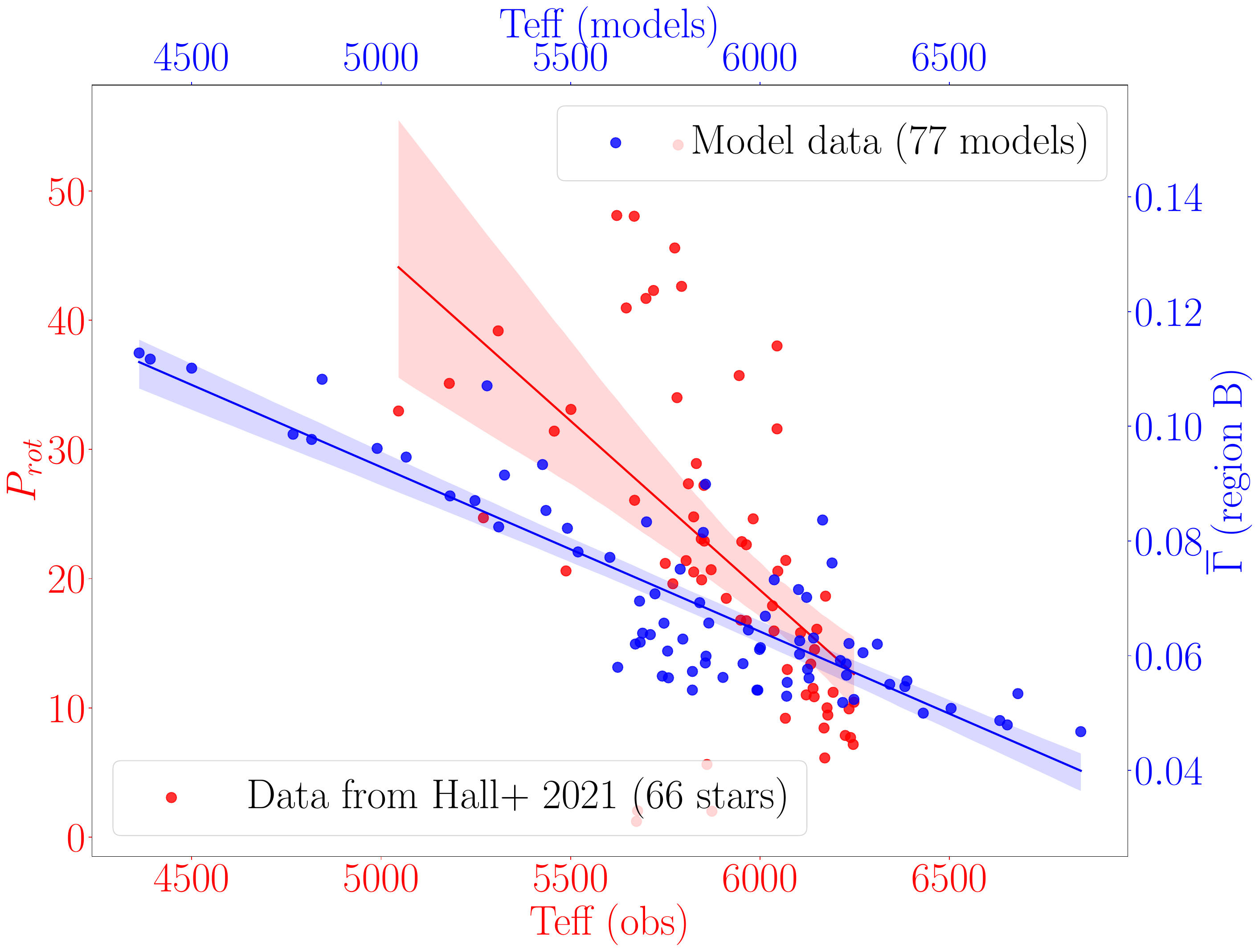}
        \includegraphics[width=6.0cm]{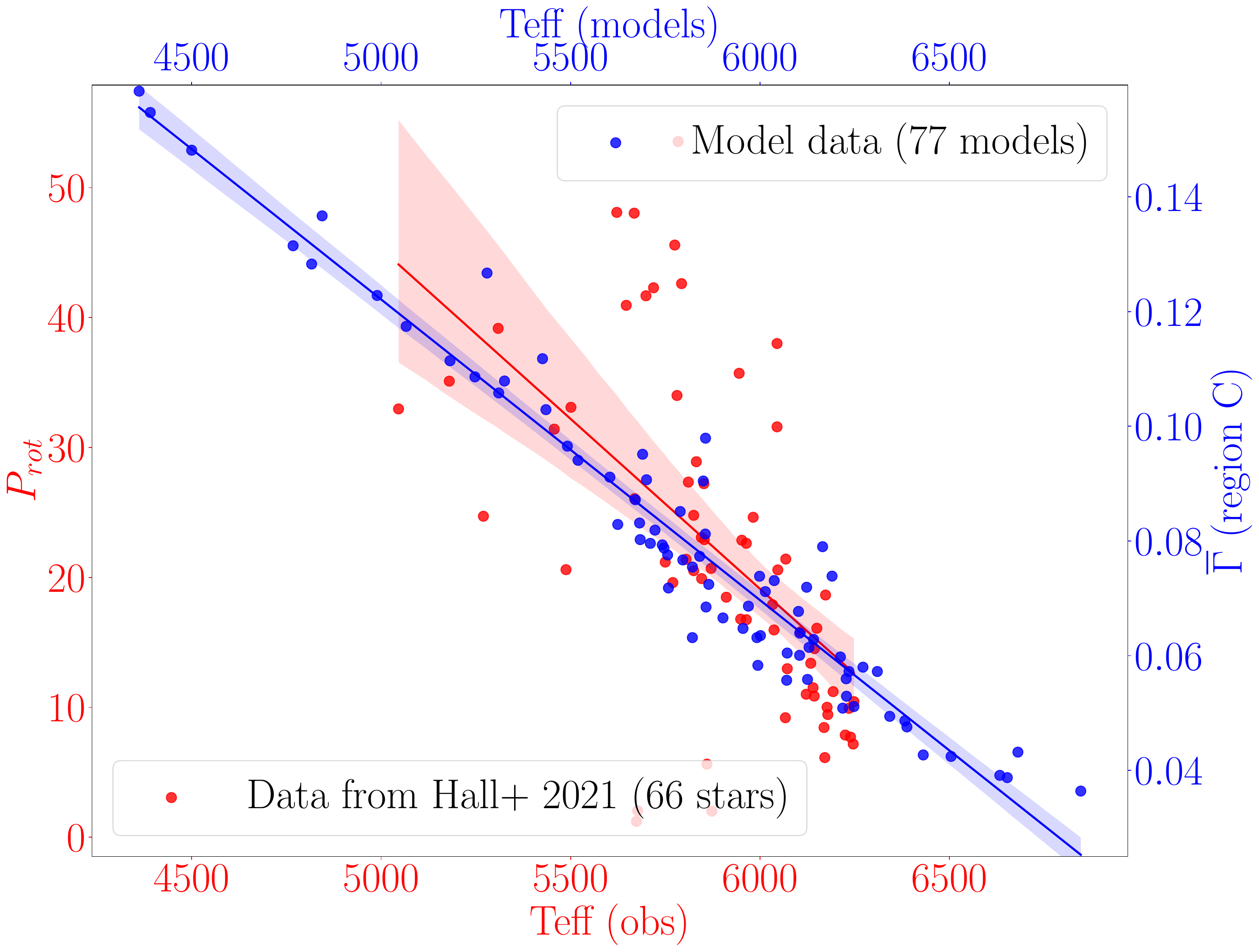}
        \caption{Rotation periods as a function of effective temperature for a group of 66  low-mass MS stars in the Kepler field (the same stars that are represented in Figure \ref{fig1} with red color, and that serve as references for the theoretical models). Red circles represent observational data.  Blue circles represent theoretical data. Specifically, they denote the $\overline{\Gamma}$ values computed for different regions within the star, as indicated on the right y-axis of each subplot. Also plotted in this figure is a linear regression model fit to each set of data. The shaded area around the regression line represents a 95\% confidence interval for the regression estimate.}
        \label{fig8}
\end{figure*}

\subsection{The electron number density, the electron degeneracy parameter, and the mean molecular weight}\label{subsec:3.4}

\begin{figure*}
        \centering
        \includegraphics[width=17cm]{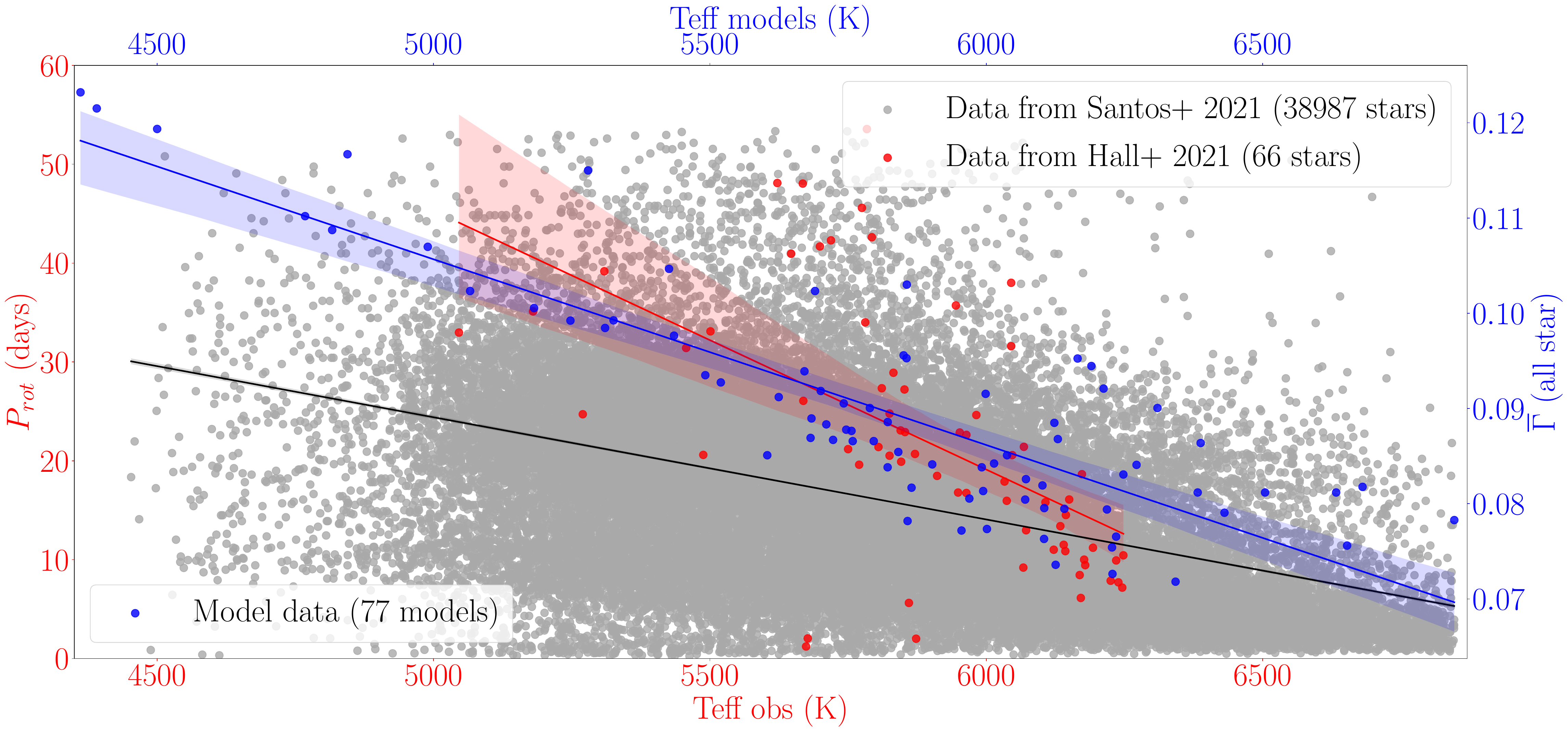}
        \caption{Rotation periods as a function of effective temperature for a group of 38,987 low-mass MS stars in the Kepler field, compared with a smaller group of 66 Kepler low-mass main-sequence stars and with theoretical data from models. Data represented in blue and red are the same data shown in Figure \ref{fig8}. In this plot, we add a third regression model fitted to the observational data, for a set of 38987 stars from the catalog by \cite{2021yCat..22550017S}.
        }
        \label{fig9}
\end{figure*}

In relation to the results of Figure \ref{fig4}, it would be useful to find out whether or not the dependence of $\overline{\Gamma}$ on the effective temperature is merely a consequence of stellar structure and evolution that can be replicated with other fundamental physical parameters. Alternatively, the dependence of $\overline{\Gamma}$ on the effective temperature may exhibit a unique character that reveals a significant connection between the interiors of stars and the observed rotational properties of low-mass MS stars.

In an attempt to decipher between these two possibilities, we also computed global mean values for the electron number density ($n_e$), the electron degeneracy parameter ($\eta$), and the mean molecular weight ($\mu$). Definitions of these quantities can be found in, for example, \citet{2013sse..book.....K}. The global mean values were calculated in exactly the same way as the global plasma coupling parameter. Namely,
\begin{equation}
                \overline{n_e} = \frac{1}{M} \int_{0}^{M} n_e \, dM(r) \, ,
                \label{eq12}
\end{equation}
\begin{equation}
        \overline{\eta} = \frac{1}{M} \int_{0}^{M} \eta \, dM(r) \, ,
        \label{eq13}
\end{equation}
and
\begin{equation}
        \overline{\mu} = \frac{1}{M} \int_{0}^{M} \mu \, dM(r) \, ,
        \label{eq14}
\end{equation}
where $dM(r)$ is again given by Equation \ref{eq11} .

The plots for the dependence of these three quantities on the effective temperature are shown in Figure \ref{fig5}. 
The global electron density dependence on the effective temperature (left panel of Figure \ref{fig5}) does not share the scaling properties of the global plasma parameter. In this case, we cannot distinguish a clear scaling of $\overline{n_e}$ with the effective temperature, as the groups of stars with masses of $1.0, 1.1, 1.2, 1.3$, and $1.4$ \, $M_\odot$ have mean electron densities that span the entire possible range of values. Similarly, the dependence of the  global
mean molecular weight on effective temperature does not exhibit the strong scaling properties of $\overline{\Gamma}$. Nevertheless, cooler stars generally have lower values of $\overline{\mu}$ than hotter stars. As is well known, the mean molecular weight is sensitive to the composition of the stellar material. We show in the next section that the global plasma parameter strongly depends on metallicity. Finally, as a consequence of the Pauli exclusion principle, electrons in the interiors of stars become degenerate. The electron degeneracy parameter, $\eta$, measures the degree of degeneracy, with larger values of $\eta$ indicating more significant degeneracy. The central panel of Figure \ref{fig5} shows the global degeneracy parameter, $\overline{\eta}$,  as a function of effective temperature. Here, unlike the other two cases discussed above ($\overline{n_e}$ and $\overline{\mu}$), we notice that $\overline{\eta}$ shares similar scaling properties with the global plasma parameter. However, the dispersion of $\overline{\eta}$ is much more pronounced than that of $\overline{\Gamma}$. From the perspective of the importance of nonideal effects in stellar interiors, particularly in the cooler upper layers of low-mass stars, this is interesting because two of the main contributors to the nonideal character of the equation of state in this region are pressure ---due to electrostatic interactions--- and electron degeneracy \citep[e.g.,][]{1968psen.book.....C}.

Regarding the question we outline at the beginning of this subsection, we can now conclude that the dependence of $\overline{\Gamma}$ on the effective temperature is not merely a consequence of stellar structure and evolution and cannot be easily replicated with other fundamental physical parameters. In the following section, we continue our investigation of the properties of $\overline{\Gamma}$, namely of its dependence on age and metallicity.

\section{Dependence of the global plasma coupling parameter on age and metallicity}\label{sec:4}

\subsection{Age}\label{subsec:4.1}

The evolution of the global plasma parameter, $\overline{\Gamma}$, throughout the entire MS lifetime is shown in Figures \ref{fig6} and \ref{fig7} for different stellar masses and metallicities. The ages at ZAMS and at TAMS for the models at Z=0.02 metallicity represented in Figures \ref{fig6} and \ref{fig7} are listed in Table \ref{tab2} . We used the central mass fraction of hydrogen to define the TAMS; in our models, the TAMS is reached when the hydrogen mass fraction drops below $ 10^{-9}$.

Figure \ref{fig6} focuses on comparing the global plasma parameter for stellar models with different masses but the same metallicity (all models in Figure \ref{fig6} were computed at Z=0.02 metallicity). This figure demonstrates that the relevance of electrostatic interactions throughout the MS lifetime ---for the same metallicity--- is entirely mass dependent. Specifically, as the mass of the star decreases, the value of the global parameter increases. This supports the expected result that as the mass of the star decreases, the Coulomb interactions between particles become increasingly significant in the thermodynamics of stellar interiors.
From Figure \ref{fig7}, we also observe that the values of the global plasma parameter decrease from ZAMS up to approximately 75\% of the MS lifetime, and then begin to increase until the end of the MS.
From the behavior of $\overline{\Gamma}$ as a function of age, and knowing that slower rotators are associated with higher values of $\overline{\Gamma}$ while rapid rotators are associated with lower values of $\overline{\Gamma}$, it is natural to infer that as the values of $\overline{\Gamma}$ decrease over approximately three-quarters of the MS lifetime, the stellar spin-down should weaken (again, because $\overline{\Gamma}$ is decreasing). Interestingly, in recent years, some observational studies \citep[e.g.,][]{2016Natur.529..181V, 2020ApJ...904..140C, 2021NatAs...5..707H} have proposed a scenario of weakened magnetic braking that deviates from the standard stellar spin-down laws. We have known about the link between the rotation rate of a star and its
age since the seminal works of \citet{1967ApJ...150..551K} and \citet{1972ApJ...171..565S}. Specifically, based on observational data, Skumanich derived that for low-mass stars, $v_{\text{eq}} \propto t^{-1/2}$, where $v_{\text{eq}}$ is the equatorial rotational velocity and $t$ is the stellar age. In the weakened magnetic braking scenario, the star deviates from the standard Skumanich law and appears to be stalled at intermediate or older ages in the MS \citep[e.g.,][]{2024ApJ...962..138S}. The underlying physical mechanism that leads to this weakened magnetic braking remains unknown. Nevertheless, we want to emphasize that the theoretical properties of the global plasma parameter, particularly its dependence on age, can be related to the weakened magnetic braking scenario. Moreover, Figure \ref{fig7} shows that in the last quarter of the MS, the global plasma parameter increases. If $\overline{\Gamma}$ can indeed be related to stellar rotation rates, this would suggest that older MS stars might resume spin down after a period of stalling. This behavior of resumed spin down at older ages was reported for stars in the Ruprecht 147 cluster by \citet{2020ApJ...904..140C}. 

\subsection{Metallicity}\label{subsec:4.2}

The effect of metallicity on the global plasma parameter is clearly marked, with high-metallicity stellar models exhibiting higher values of $\overline{\Gamma}$.  For all the different categories of stellar masses, the metallicity has a large impact on the global plasma parameter throughout the entire MS evolution. Figure \ref{fig7} unambiguously demonstrates the impact of metallicity on the global plasma parameter. This means that, during the MS lifetimes of all the represented stellar models, the higher the metallicity, the greater the significance of electrostatic effects in these stellar interiors.
The relationships between rotation, magnetic activity, and metallicity are still poorly understood. However, some recent studies have unveiled several insights into how metallicity might affect stellar rotation and activity \citep{2020MNRAS.499.3481A, 2020ApJ...898...76S, 2021ApJ...912..127S, 2022ApJ...930....7A}. Specifically, the study by \citet{2020MNRAS.499.3481A}, which considered thousands of Kepler stars with masses ranging from $0.85$ to $1.3 \, M_\odot$ ---nearly the mass range addressed in the present study— found a correlation between metallicity and rotation: metal-rich stars tend to rotate slower than metal-poor stars. Another very recent study identified a link between slow rotation and high metallicity values for a specific group of Kepler stars \citep{2023A&A...672A..56S}. Here again, we can link the properties of the global plasma parameter with the observed rotation rates, that is, the higher the metallicity of the stellar model, the higher the value of the global plasma parameter, which in turn can be related to lower rotation rates.

\section{Electrostatic effects and the stellar rotation rates of low-mass stars on the MS}\label{sec:5}

\medskip

It is well known that one-dimensional stellar models include several approximations that allow us to solve the set of differential equations plus the corresponding boundary conditions. These approximations are particularly impactful if rotation and magnetic fields are not considered, because spherical symmetry is conserved. Nonetheless, it is still possible to obtain revealing insights by looking at the physical ingredients that can be linked to rotation and magnetism. One of these ingredients is the relevance of electrostatic interactions, because in a stellar plasma, all particles experience the Lorentz force, and the Lorentz force can in turn be related to mechanisms of angular momentum transport in stellar interiors \citep[e.g.,][]{2019ARA&A..57...35A}

The discovery of the solar wind by \citet{1958ApJ...128..664P} boosted the studies of stellar magnetized winds. It is thought that these winds exert a breaking torque that removes angular momentum from the stars \citep[e.g.,][]{1962AnAp...25...18S, 1967ApJ...148..217W, 1968MNRAS.138..359M, 1988ApJ...333..236K, 1997ApJ...480..303K, 2000ApJ...534..335S, 2011MNRAS.416..447S}. Taking into consideration all the accumulated empirical data and also the theoretical predictions, stellar rotation appears to depend strongly on mass, age, and metallicity. In the previous sections of this work we study the properties of a global plasma parameter, $\overline{\Gamma}$, which allows us to compare the importance of electrostatic effects among a group of stellar models. We also explore some connections of this global plasma parameter with the observed patterns for rotation in the MS. In the present section, we again use our set of theoretical models described in Section \ref{sec:2} to further investigate how the properties of the global plasma parameter relate to the rotational observational data for low-mass MS stars.

The sample of stars on which we base our models (see Figure \ref{fig2}) is a sample of stars with asteroseismically measured rotation periods \citep{2021NatAs...5..707H}. This means that the values obtained for the rotation periods do not depend on the starspot variations. To investigate further whether the electrostatic properties of stellar interiors correlate with their observed rotational trends, we plot the observational rotation periods versus the (also observed) effective temperature, superimposed with the variation of the global plasma parameter as a function of effective temperature obtained from our models. These plots are shown in Figure \ref{fig8} , where the global plasma parameter is computed for different regions within the interiors of the stellar models. Red circles stand for observational data, whereas blue circles indicate theoretical data from models. We note that the behavior of $\overline{\Gamma}$ as a function of effective temperature unexpectedly follows a scaling, with the slope of this line being very similar to the scaling observed in the relationship between rotation periods and effective temperature. Moreover, the outer layers of the stars appear to be related to the slope of the linear scaling, whereas the core region of the star introduces a dispersion around the global trend. The fact that the outer layers have a marked contribution should not be a surprise as the base of the convective zone, located in the upper 30\% of the stellar radius, is thought to be responsible for a dynamo mechanism of magnetic field amplification \citep[e.g.,][]{2009SoPh..257....1L,2012MNRAS.422.1709P, 2014SSRv..186..535L}. Moreover, \citet{2019MNRAS.488.1558B} also found a correlation between a structural feature occurring in the outer convective layers and the rotational behaviors of a group of Kepler stars.

To better understand these trends, we fitted a linear regression model to the data using the least squares method. The fits are also represented in Figure \ref{fig8}. Because the observational data are limited to 66 low-mass MS stars in the Kepler field, we aim to investigate whether the similarity between the two trends (the observational trend and the theoretical trend shown in Figure \ref{fig8} ) is a feature of this particular group of Kepler stars, or if the same correlation can be obtained when considering other groups of stars. For this purpose, we used the Kepler sample published in \citet{2021ApJS..255...17S} to compare with our theoretical model results. From this large catalog of observational data, we selected all stars within the $T_{\text{eff}}$ window of our models. This choice results in a dataset of 38987 Kepler stars. In order to compare all of the groups of stars, that is, the 66 stars from \citet{2021NatAs...5..707H}, the large sample from \citet{2021ApJS..255...17S}, and our theoretical results for the global plasma parameter, we normalized all three datasets to their maximum values. The results for the linear regression model fits for these three datasets are shown in Figure \ref{fig9}, and the regression outputs are given in Table \ref{tab3}. The results evidenced by the regression models in Figure \ref{fig9} reveal a strong correlation between the properties of $\overline{\Gamma}$ and the observed rotational trends. Based on our analysis of the results for the regression lines, we conclude that the correlation between the theoretical trend of the global plasma parameter with the effective temperature, and the observational trend of the stellar rotation period and the effective temperature, is not a consequence of using a small dataset (with 66 Kepler stars). This correlation between the two trends is even closer when we use a large observational dataset of Kepler stars.


\begin{table}
        \caption{Regression outputs}              
        \label{tab3}      
        \centering                                      
        \begin{tabular}{c c c c}          
                \hline\hline                        
                Data & Slope & Intercept & R-squared \\    
                \hline                                   
                Hall+ & $-4.88 \times 10^{-4}$ & 3.29 & 0.312\\      
                Santos + & $-1.94 \times 10^{-4}$ &  1.43 &  0.232\\
                Models & $-1.58 \times 10^{-4}$ & 1.64 & 0.731\\
                \hline                                             
        \end{tabular}
\tablefoot{
The observational datasets used in these regression models were sourced from: \citet{2021NatAs...5..707H} (Hall+) and \citet{2021ApJS..255...17S} (Santos+).
}
\end{table}

\section{Conclusions}
\label{sec:6}

The connections between the internal thermodynamic behavior of the stellar plasma and stellar rotation represent an understudied research field. From the observational data, we know that rotation is an intricate function of mass, age, and metallicity. In a previous work, we demonstrated that Coulomb interactions can significantly impact the sound-speed gradient in the outer layers of lower-mass stars \citep{2021MNRAS.507.5747B}. In the present paper, we focus on describing the electrostatic effects comprehensively across a set of low-mass theoretical stellar models, and correlate these effects with the observed main rotational trends for low-mass stars on the MS.

 By studying the electrostatic properties of stellar interiors across a set of theoretical MS stellar models with masses ranging from $0.7$ to $1.4 \, M_\odot$, we have discovered that this type of microphysics in stellar interiors can be directly linked to observed rotational behaviors in stars within this mass range. Our main conclusions can be summarized as follows:

1. The linear scaling of the global plasma parameter with effective temperature shows a very high level of similarity to the linear scaling of rotation periods with effective temperature. Slow rotators can be linked to stars where electrostatic effects are more significant, whereas rapid rotators can be linked to stars where these effects are less significant. This similarity is not merely a consequence of stellar evolution, as it cannot be replicated for other stellar parameters.

2. The significance of electrostatic effects, as measured by $\overline{\Gamma}$, decreases in stellar interiors as stars age on the MS; it reaches a minimal value at approximately 75\% of the MS lifetime and then starts to increase again. This behavior can be linked to an observational phenomenon known as weakened magnetic braking \citep[e.g.,][]{2016Natur.529..181V, 2020ApJ...904..140C}.

3. High-metallicity stellar models exhibit higher values of the global plasma parameter compared to low-metallicity models, indicating that metallicity significantly impacts electrostatic interactions within stellar interiors. The observational data linking rotation and metallicity is limited. Some studies suggest that high-metallicity stars rotate more slowly than low-metallicity stars \citep[e.g.,][]{2020MNRAS.499.3481A}. The properties of the global plasma parameter can be related to this scenario.

In this initial study, we use a set of nonrotating low-mass stellar models to explore the relationships between the electrostatic properties of stellar interiors and the observed rotation rates for stars in this mass range. However, given the impact that metallicity has on these electrostatic properties (see subsection \ref{subsec:4.2}) and previous studies with rotating models that demonstrated the influence of metallicity on internal angular momentum transport and the rotation periods of low-mass stars \citep[e.g.,][]{2016A&A...587A.105A, 2019A&A...631A..77A}, we believe that further investigation of electrostatic properties using rotating models will significantly improve our knowledge regarding this topic. 
                
It is also well known that the observed rotation rates of red giant cores are in disagreement with theoretical predictions. In stellar interiors, the fluid constituting the stellar material experiences Lorentz forces, which are known to play an important role in angular momentum transport. For example, the torque generated by the so-called Tayler-Spruit dynamo \citep{2002A&A...381..923S} could represent an important step forward in understanding this disagreement. The implementation of the Tayler-Spruit dynamo in the MESA stellar evolution code by \citet{2014ApJ...788...93C} for low-mass stars showed that models with this implementation are in better agreement with the observed core rotation rates. Subsequently, a revised formulation of the Tayler-Spruit dynamo, known as the Fuller formalism \citep{2019MNRAS.485.3661F}, led to an even more efficient transport of angular momentum in the cores of red giants and consequently to better agreement with observational data. Moreover, a mechanism of angular momentum transport based on the Tayler instability was also successful in explaining the uniform rotation observed in the solar radiative zone \citep{2005A&A...440L...9E}. Although the Tayler-Spruit mechanism is being used in stellar evolution codes, it still has physical deficiencies, as  pointed out in several works \citep[e.g.,][]{2007A&A...474..145Z, 2006A&A...449..451B, 2019ApJ...881...66G}. Therefore, understanding the properties of electrostatic interactions in stellar interiors that can be related to the transport of angular momentum by magnetic fields \citep[e.g.,][]{1999ApJ...519..911M, 2000ASPC..198..329M} could significantly improve our comprehension of the mechanisms behind the redistribution of angular momentum  in low-mass stars.

\begin{acknowledgements}
We thank the anonymous referee for the valuable comments and remarks, which have significantly improved the clarity and accuracy of the manuscript. 
The authors A. Brito and I. Lopes also thank the Funda\c c\~ao para a Ci\^encia e Tecnologia (FCT), Portugal for the financial support to the Center for Astrophysics and Gravitation--CENTRA, Instituto Superior Tecnico, Universidade de Lisboa, through the Project No. UIDB/00099/2020. and grant No. PTDC/FISAST/ 28920/2017.
\end{acknowledgements}

%
%

\bibliographystyle{aa} 
\bibliography{bib_file} 

\begin{thebibliography}{92}
\expandafter\ifx\csname natexlab\endcsname\relax\def\natexlab#1{#1}\fi

\bibitem[{{Aerts} {et~al.}(2019){Aerts}, {Mathis}, \&
  {Rogers}}]{2019ARA&A..57...35A}
{Aerts}, C., {Mathis}, S., \& {Rogers}, T.~M. 2019, \araa, 57, 35

\bibitem[{{Amard} {et~al.}(2016){Amard}, {Palacios}, {Charbonnel}, {Gallet}, \&
  {Bouvier}}]{2016A&A...587A.105A}
{Amard}, L., {Palacios}, A., {Charbonnel}, C., {Gallet}, F., \& {Bouvier}, J.
  2016, \aap, 587, A105

\bibitem[{{Amard} {et~al.}(2019){Amard}, {Palacios}, {Charbonnel}, {Gallet},
  {Georgy}, {Lagarde}, \& {Siess}}]{2019A&A...631A..77A}
{Amard}, L., {Palacios}, A., {Charbonnel}, C., {et~al.} 2019, \aap, 631, A77

\bibitem[{{Amard} {et~al.}(2020){Amard}, {Roquette}, \&
  {Matt}}]{2020MNRAS.499.3481A}
{Amard}, L., {Roquette}, J., \& {Matt}, S.~P. 2020, \mnras, 499, 3481

\bibitem[{{Asplund} {et~al.}(2009){Asplund}, {Grevesse}, {Sauval}, \&
  {Scott}}]{2009ARA&A..47..481A}
{Asplund}, M., {Grevesse}, N., {Sauval}, A.~J., \& {Scott}, P. 2009, \araa, 47,
  481

\bibitem[{{Avallone} {et~al.}(2022){Avallone}, {Tayar}, {van Saders}, {Berger},
  {Claytor}, {Beaton}, {Teske}, {Godoy-Rivera}, \& {Pan}}]{2022ApJ...930....7A}
{Avallone}, E.~A., {Tayar}, J.~N., {van Saders}, J.~L., {et~al.} 2022, \apj,
  930, 7

\bibitem[{{Bahcall} {et~al.}(2004){Bahcall}, {Serenelli}, \&
  {Pinsonneault}}]{2004ApJ...614..464B}
{Bahcall}, J.~N., {Serenelli}, A.~M., \& {Pinsonneault}, M. 2004, \apj, 614,
  464

\bibitem[{{Barnes}(2003)}]{2003ApJ...586..464B}
{Barnes}, S.~A. 2003, \apj, 586, 464

\bibitem[{{Barnes} \& {Kim}(2010)}]{2010ApJ...721..675B}
{Barnes}, S.~A. \& {Kim}, Y.-C. 2010, \apj, 721, 675

\bibitem[{{Basu} \& {Antia}(2008)}]{2008PhR...457..217B}
{Basu}, S. \& {Antia}, H.~M. 2008, \physrep, 457, 217

\bibitem[{{B{\"o}hm-Vitense}(1958)}]{1958ZA.....46..108B}
{B{\"o}hm-Vitense}, E. 1958, \zap, 46, 108

\bibitem[{{B{\"o}hm-Vitense}(2007)}]{2007ApJ...657..486B}
{B{\"o}hm-Vitense}, E. 2007, \apj, 657, 486

\bibitem[{{Braithwaite}(2006)}]{2006A&A...449..451B}
{Braithwaite}, J. 2006, \aap, 449, 451

\bibitem[{{Brito} \& {Lopes}(2019)}]{2019MNRAS.488.1558B}
{Brito}, A. \& {Lopes}, I. 2019, \mnras, 488, 1558

\bibitem[{{Brito} \& {Lopes}(2021)}]{2021MNRAS.507.5747B}
{Brito}, A. \& {Lopes}, I. 2021, \mnras, 507, 5747

\bibitem[{{Cantiello} {et~al.}(2014){Cantiello}, {Mankovich}, {Bildsten},
  {Christensen-Dalsgaard}, \& {Paxton}}]{2014ApJ...788...93C}
{Cantiello}, M., {Mankovich}, C., {Bildsten}, L., {Christensen-Dalsgaard}, J.,
  \& {Paxton}, B. 2014, \apj, 788, 93

\bibitem[{{Chen}(2016)}]{2016ippc.book.....C}
{Chen}, F.~F. 2016, {Introduction to Plasma Physics and Controlled Fusion}

\bibitem[{{Christensen-Dalsgaard}(2021)}]{2021LRSP...18....2C}
{Christensen-Dalsgaard}, J. 2021, Living Reviews in Solar Physics, 18, 2

\bibitem[{{Chugunov} {et~al.}(2007){Chugunov}, {Dewitt}, \&
  {Yakovlev}}]{Chugunov2007}
{Chugunov}, A.~I., {Dewitt}, H.~E., \& {Yakovlev}, D.~G. 2007, \prd, 76, 025028

\bibitem[{{Clayton}(1968)}]{1968psen.book.....C}
{Clayton}, D.~D. 1968, {Principles of stellar evolution and nucleosynthesis}

\bibitem[{{Curtis} {et~al.}(2020){Curtis}, {Ag{\"u}eros}, {Matt}, {Covey},
  {Douglas}, {Angus}, {Saar}, {Cody}, {Vanderburg}, {Law}, {Kraus}, {Latham},
  {Baranec}, {Riddle}, {Ziegler}, {Lund}, {Torres}, {Meibom}, {Aguirre}, \&
  {Wright}}]{2020ApJ...904..140C}
{Curtis}, J.~L., {Ag{\"u}eros}, M.~A., {Matt}, S.~P., {et~al.} 2020, \apj, 904,
  140

\bibitem[{{Cyburt} {et~al.}(2010){Cyburt}, {Amthor}, {Ferguson}, {Meisel},
  {Smith}, {Warren}, {Heger}, {Hoffman}, {Rauscher}, {Sakharuk}, {Schatz},
  {Thielemann}, \& {Wiescher}}]{Cyburt2010}
{Cyburt}, R.~H., {Amthor}, A.~M., {Ferguson}, R., {et~al.} 2010, \apjs, 189,
  240

\bibitem[{{Deal} {et~al.}(2018){Deal}, {Alecian}, {Lebreton}, {Goupil},
  {Marques}, {LeBlanc}, {Morel}, \& {Pichon}}]{2018A&A...618A..10D}
{Deal}, M., {Alecian}, G., {Lebreton}, Y., {et~al.} 2018, \aap, 618, A10

\bibitem[{Debye \& Hückel(1923)}]{Debye_1923}
Debye, P. \& Hückel, E. 1923, Physikalische Zeitschrift, 24, 305

\bibitem[{{do Nascimento} {et~al.}(2014){do Nascimento}, {Garc{\'\i}a},
  {Mathur}, {Anthony}, {Barnes}, {Meibom}, {da Costa}, {Castro}, {Salabert}, \&
  {Ceillier}}]{2014ApJ...790L..23D}
{do Nascimento}, J.~D., J., {Garc{\'\i}a}, R.~A., {Mathur}, S., {et~al.} 2014,
  \apjl, 790, L23

\bibitem[{{Eggenberger} {et~al.}(2005){Eggenberger}, {Maeder}, \&
  {Meynet}}]{2005A&A...440L...9E}
{Eggenberger}, P., {Maeder}, A., \& {Meynet}, G. 2005, \aap, 440, L9

\bibitem[{{Epstein} \& {Pinsonneault}(2014)}]{2014ApJ...780..159E}
{Epstein}, C.~R. \& {Pinsonneault}, M.~H. 2014, \apj, 780, 159

\bibitem[{{Ferguson} {et~al.}(2005){Ferguson}, {Alexander}, {Allard}, {Barman},
  {Bodnarik}, {Hauschildt}, {Heffner-Wong}, \& {Tamanai}}]{2005ApJ...623..585F}
{Ferguson}, J.~W., {Alexander}, D.~R., {Allard}, F., {et~al.} 2005, \apj, 623,
  585

\bibitem[{{Fuller} {et~al.}(2019){Fuller}, {Piro}, \&
  {Jermyn}}]{2019MNRAS.485.3661F}
{Fuller}, J., {Piro}, A.~L., \& {Jermyn}, A.~S. 2019, \mnras, 485, 3661

\bibitem[{{Garc{\'\i}a} {et~al.}(2014){Garc{\'\i}a}, {Ceillier}, {Salabert},
  {Mathur}, {van Saders}, {Pinsonneault}, {Ballot}, {Beck}, {Bloemen},
  {Campante}, {Davies}, {do Nascimento}, {Mathis}, {Metcalfe}, {Nielsen},
  {Su{\'a}rez}, {Chaplin}, {Jim{\'e}nez}, \& {Karoff}}]{2014A&A...572A..34G}
{Garc{\'\i}a}, R.~A., {Ceillier}, T., {Salabert}, D., {et~al.} 2014, \aap, 572,
  A34

\bibitem[{{Goldstein} {et~al.}(2019){Goldstein}, {Townsend}, \&
  {Zweibel}}]{2019ApJ...881...66G}
{Goldstein}, J., {Townsend}, R.~H.~D., \& {Zweibel}, E.~G. 2019, \apj, 881, 66

\bibitem[{{Hall}(1991)}]{1991LNP...380..353H}
{Hall}, D.~S. 1991, in IAU Colloq. 130: The Sun and Cool Stars. Activity,
  Magnetism, Dynamos, ed. I.~{Tuominen}, D.~{Moss}, \& G.~{R{\"u}diger}, Vol.
  380, 353

\bibitem[{{Hall} {et~al.}(2021){Hall}, {Davies}, {van Saders}, {Nielsen},
  {Lund}, {Chaplin}, {Garc{\'\i}a}, {Amard}, {Breimann}, {Khan}, {See}, \&
  {Tayar}}]{2021NatAs...5..707H}
{Hall}, O.~J., {Davies}, G.~R., {van Saders}, J., {et~al.} 2021, Nature
  Astronomy, 5, 707

\bibitem[{{Hempelmann} {et~al.}(1995){Hempelmann}, {Schmitt}, {Schultz},
  {Ruediger}, \& {Stepien}}]{1995A&A...294..515H}
{Hempelmann}, A., {Schmitt}, J.~H.~M.~M., {Schultz}, M., {Ruediger}, G., \&
  {Stepien}, K. 1995, \aap, 294, 515

\bibitem[{{Iglesias} \& {Rogers}(1993)}]{Iglesias1993}
{Iglesias}, C.~A. \& {Rogers}, F.~J. 1993, \apj, 412, 752

\bibitem[{{Iglesias} \& {Rogers}(1996)}]{1996ApJ...464..943I}
{Iglesias}, C.~A. \& {Rogers}, F.~J. 1996, \apj, 464, 943

\bibitem[{{Irwin}(2004)}]{Irwin2004}
{Irwin}, A.~W. 2004, The FreeEOS Code for Calculating the Equation of State for
  Stellar Interiors

\bibitem[{{Jermyn} {et~al.}(2023){Jermyn}, {Bauer}, {Schwab}, {Farmer}, {Ball},
  {Bellinger}, {Dotter}, {Joyce}, {Marchant}, {Mombarg}, {Wolf}, {Sunny Wong},
  {Cinquegrana}, {Farrell}, {Smolec}, {Thoul}, {Cantiello}, {Herwig}, {Toloza},
  {Bildsten}, {Townsend}, \& {Timmes}}]{2023ApJS..265...15J}
{Jermyn}, A.~S., {Bauer}, E.~B., {Schwab}, J., {et~al.} 2023, \apjs, 265, 15

\bibitem[{{Jermyn} {et~al.}(2021){Jermyn}, {Schwab}, {Bauer}, {Timmes}, \&
  {Potekhin}}]{Jermyn2021}
{Jermyn}, A.~S., {Schwab}, J., {Bauer}, E., {Timmes}, F.~X., \& {Potekhin},
  A.~Y. 2021, \apj, 913, 72

\bibitem[{{Kawaler}(1988)}]{1988ApJ...333..236K}
{Kawaler}, S.~D. 1988, \apj, 333, 236

\bibitem[{{Kippenhahn} {et~al.}(2013){Kippenhahn}, {Weigert}, \&
  {Weiss}}]{2013sse..book.....K}
{Kippenhahn}, R., {Weigert}, A., \& {Weiss}, A. 2013, {Stellar Structure and
  Evolution}

\bibitem[{{Kraft}(1967)}]{1967ApJ...150..551K}
{Kraft}, R.~P. 1967, \apj, 150, 551

\bibitem[{{Krishnamurthi} {et~al.}(1997){Krishnamurthi}, {Pinsonneault},
  {Barnes}, \& {Sofia}}]{1997ApJ...480..303K}
{Krishnamurthi}, A., {Pinsonneault}, M.~H., {Barnes}, S., \& {Sofia}, S. 1997,
  \apj, 480, 303

\bibitem[{{Lopes} \& {Passos}(2009)}]{2009SoPh..257....1L}
{Lopes}, I. \& {Passos}, D. 2009, \solphys, 257, 1

\bibitem[{{Lopes} {et~al.}(2014){Lopes}, {Passos}, {Nagy}, \&
  {Petrovay}}]{2014SSRv..186..535L}
{Lopes}, I., {Passos}, D., {Nagy}, M., \& {Petrovay}, K. 2014, \ssr, 186, 535

\bibitem[{{MacGregor}(2000)}]{2000ASPC..198..329M}
{MacGregor}, K.~B. 2000, in Astronomical Society of the Pacific Conference
  Series, Vol. 198, Stellar Clusters and Associations: Convection, Rotation,
  and Dynamos, ed. R.~{Pallavicini}, G.~{Micela}, \& S.~{Sciortino}, 329

\bibitem[{{MacGregor} \& {Charbonneau}(1999)}]{1999ApJ...519..911M}
{MacGregor}, K.~B. \& {Charbonneau}, P. 1999, \apj, 519, 911

\bibitem[{{Maeder}(2009)}]{2009pfer.book.....M}
{Maeder}, A. 2009, {Physics, Formation and Evolution of Rotating Stars}

\bibitem[{{Mamajek} \& {Hillenbrand}(2008)}]{2008ApJ...687.1264M}
{Mamajek}, E.~E. \& {Hillenbrand}, L.~A. 2008, \apj, 687, 1264

\bibitem[{{Marsden} {et~al.}(2014){Marsden}, {Petit}, {Jeffers}, {Morin},
  {Fares}, {Reiners}, {do Nascimento}, {Auri{\`e}re}, {Bouvier}, {Carter},
  {Catala}, {Dintrans}, {Donati}, {Gastine}, {Jardine}, {Konstantinova-Antova},
  {Lanoux}, {Ligni{\`e}res}, {Morgenthaler}, {Ram{\`\i}rez-V{\`e}lez},
  {Th{\'e}ado}, {Van Grootel}, \& {BCool Collaboration}}]{2014MNRAS.444.3517M}
{Marsden}, S.~C., {Petit}, P., {Jeffers}, S.~V., {et~al.} 2014, \mnras, 444,
  3517

\bibitem[{{Mathur} {et~al.}(2023){Mathur}, {Claytor}, {Santos}, {Garc{\'\i}a},
  {Amard}, {Bugnet}, {Corsaro}, {Bonanno}, {Breton}, {Godoy-Rivera},
  {Pinsonneault}, \& {van Saders}}]{2023ApJ...952..131M}
{Mathur}, S., {Claytor}, Z.~R., {Santos}, {\^A}. R.~G., {et~al.} 2023, \apj,
  952, 131

\bibitem[{{Mestel}(1968)}]{1968MNRAS.138..359M}
{Mestel}, L. 1968, \mnras, 138, 359

\bibitem[{{Mestel}(1999)}]{1999stma.book.....M}
{Mestel}, L. 1999, {Stellar magnetism}

\bibitem[{{Moedas} {et~al.}(2022){Moedas}, {Deal}, {Bossini}, \&
  {Campilho}}]{2022A&A...666A..43M}
{Moedas}, N., {Deal}, M., {Bossini}, D., \& {Campilho}, B. 2022, \aap, 666, A43

\bibitem[{{Nsamba} {et~al.}(2018){Nsamba}, {Campante}, {Monteiro}, {Cunha},
  {Rendle}, {Reese}, \& {Verma}}]{2018MNRAS.477.5052N}
{Nsamba}, B., {Campante}, T.~L., {Monteiro}, M.~J.~P.~F.~G., {et~al.} 2018,
  \mnras, 477, 5052

\bibitem[{{Ol{\'a}h} {et~al.}(2016){Ol{\'a}h}, {K{\H{o}}v{\'a}ri}, {Petrovay},
  {Soon}, {Baliunas}, {Koll{\'a}th}, \& {Vida}}]{2016A&A...590A.133O}
{Ol{\'a}h}, K., {K{\H{o}}v{\'a}ri}, Z., {Petrovay}, K., {et~al.} 2016, \aap,
  590, A133

\bibitem[{{Paquette} {et~al.}(1986){Paquette}, {Pelletier}, {Fontaine}, \&
  {Michaud}}]{1986ApJS...61..177P}
{Paquette}, C., {Pelletier}, C., {Fontaine}, G., \& {Michaud}, G. 1986, \apjs,
  61, 177

\bibitem[{{Parker}(1958)}]{1958ApJ...128..664P}
{Parker}, E.~N. 1958, \apj, 128, 664

\bibitem[{{Passos} \& {Lopes}(2012)}]{2012MNRAS.422.1709P}
{Passos}, D. \& {Lopes}, I. 2012, \mnras, 422, 1709

\bibitem[{{Paxton} {et~al.}(2011){Paxton}, {Bildsten}, {Dotter}, {Herwig},
  {Lesaffre}, \& {Timmes}}]{Paxton2011}
{Paxton}, B., {Bildsten}, L., {Dotter}, A., {et~al.} 2011, \apjs, 192, 3

\bibitem[{{Paxton} {et~al.}(2013){Paxton}, {Cantiello}, {Arras}, {Bildsten},
  {Brown}, {Dotter}, {Mankovich}, {Montgomery}, {Stello}, {Timmes}, \&
  {Townsend}}]{Paxton2013}
{Paxton}, B., {Cantiello}, M., {Arras}, P., {et~al.} 2013, \apjs, 208, 4

\bibitem[{{Paxton} {et~al.}(2015){Paxton}, {Marchant}, {Schwab}, {Bauer},
  {Bildsten}, {Cantiello}, {Dessart}, {Farmer}, {Hu}, {Langer}, {Townsend},
  {Townsley}, \& {Timmes}}]{Paxton2015}
{Paxton}, B., {Marchant}, P., {Schwab}, J., {et~al.} 2015, \apjs, 220, 15

\bibitem[{{Paxton} {et~al.}(2018){Paxton}, {Schwab}, {Bauer}, {Bildsten},
  {Blinnikov}, {Duffell}, {Farmer}, {Goldberg}, {Marchant}, {Sorokina},
  {Thoul}, {Townsend}, \& {Timmes}}]{Paxton2018}
{Paxton}, B., {Schwab}, J., {Bauer}, E.~B., {et~al.} 2018, \apjs, 234, 34

\bibitem[{{Paxton} {et~al.}(2019){Paxton}, {Smolec}, {Schwab}, {Gautschy},
  {Bildsten}, {Cantiello}, {Dotter}, {Farmer}, {Goldberg}, {Jermyn}, {Kanbur},
  {Marchant}, {Thoul}, {Townsend}, {Wolf}, {Zhang}, \& {Timmes}}]{Paxton2019}
{Paxton}, B., {Smolec}, R., {Schwab}, J., {et~al.} 2019, \apjs, 243, 10

\bibitem[{{Potekhin} \& {Chabrier}(2010)}]{Potekhin2010}
{Potekhin}, A.~Y. \& {Chabrier}, G. 2010, Contributions to Plasma Physics, 50,
  82

\bibitem[{{Potekhin} {et~al.}(2009){Potekhin}, {Chabrier}, \&
  {Rogers}}]{2009PhRvE..79a6411P}
{Potekhin}, A.~Y., {Chabrier}, G., \& {Rogers}, F.~J. 2009, \pre, 79, 016411

\bibitem[{{Rebassa-Mansergas} {et~al.}(2023){Rebassa-Mansergas}, {Maldonado},
  {Raddi}, {Torres}, {Hoskin}, {Cunningham}, {Hollands}, {Ren}, {G{\"a}nsicke},
  {Tremblay}, \& {Camisassa}}]{2023MNRAS.526.4787R}
{Rebassa-Mansergas}, A., {Maldonado}, J., {Raddi}, R., {et~al.} 2023, \mnras,
  526, 4787

\bibitem[{{Rogers} \& {Nayfonov}(2002)}]{Rogers2002}
{Rogers}, F.~J. \& {Nayfonov}, A. 2002, \apj, 576, 1064

\bibitem[{{Rose}(1998)}]{1998asa..book.....R}
{Rose}, W.~K. 1998, {Advanced Stellar Astrophysics}

\bibitem[{{Santos} {et~al.}(2021{\natexlab{a}}){Santos}, {Breton}, {Mathur}, \&
  {Garc{\'\i}a}}]{2021ApJS..255...17S}
{Santos}, A.~R.~G., {Breton}, S.~N., {Mathur}, S., \& {Garc{\'\i}a}, R.~A.
  2021{\natexlab{a}}, \apjs, 255, 17

\bibitem[{{Santos} {et~al.}(2021{\natexlab{b}}){Santos}, {Breton}, {Mathur}, \&
  {Garcia}}]{2021yCat..22550017S}
{Santos}, A.~R.~G., {Breton}, S.~N., {Mathur}, S., \& {Garcia}, R.~A.
  2021{\natexlab{b}}, {VizieR Online Data Catalog: Surface rotation \& activity
  for Kepler stars. II. (Santos+, 2021)}, VizieR On-line Data Catalog:
  J/ApJS/255/17. Originally published in: 2021ApJS..255...17S

\bibitem[{{Santos} {et~al.}(2023){Santos}, {Mathur}, {Garc{\'\i}a},
  {Broomhall}, {Egeland}, {Jim{\'e}nez}, {Godoy-Rivera}, {Breton}, {Claytor},
  {Metcalfe}, {Cunha}, \& {Amard}}]{2023A&A...672A..56S}
{Santos}, A.~R.~G., {Mathur}, S., {Garc{\'\i}a}, R.~A., {et~al.} 2023, \aap,
  672, A56

\bibitem[{{Saumon} {et~al.}(1995){Saumon}, {Chabrier}, \& {van
  Horn}}]{Saumon1995}
{Saumon}, D., {Chabrier}, G., \& {van Horn}, H.~M. 1995, \apjs, 99, 713

\bibitem[{{Saunders} {et~al.}(2024){Saunders}, {van Saders}, {Lyttle},
  {Metcalfe}, {Li}, {Davies}, {Hall}, {Ball}, {Townsend}, {Creevey}, \&
  {Dodds}}]{2024ApJ...962..138S}
{Saunders}, N., {van Saders}, J.~L., {Lyttle}, A.~J., {et~al.} 2024, \apj, 962,
  138

\bibitem[{{Schatzman}(1962)}]{1962AnAp...25...18S}
{Schatzman}, E. 1962, Annales d'Astrophysique, 25, 18

\bibitem[{{See} {et~al.}(2021){See}, {Roquette}, {Amard}, \&
  {Matt}}]{2021ApJ...912..127S}
{See}, V., {Roquette}, J., {Amard}, L., \& {Matt}, S.~P. 2021, \apj, 912, 127

\bibitem[{{Sills} {et~al.}(2000){Sills}, {Pinsonneault}, \&
  {Terndrup}}]{2000ApJ...534..335S}
{Sills}, A., {Pinsonneault}, M.~H., \& {Terndrup}, D.~M. 2000, \apj, 534, 335

\bibitem[{{Simonian} {et~al.}(2020){Simonian}, {Pinsonneault}, {Terndrup}, \&
  {van Saders}}]{2020ApJ...898...76S}
{Simonian}, G. V.~A., {Pinsonneault}, M.~H., {Terndrup}, D.~M., \& {van
  Saders}, J.~L. 2020, \apj, 898, 76

\bibitem[{{Skumanich}(1972)}]{1972ApJ...171..565S}
{Skumanich}, A. 1972, \apj, 171, 565

\bibitem[{{Spada} \& {Lanzafame}(2020)}]{2020A&A...636A..76S}
{Spada}, F. \& {Lanzafame}, A.~C. 2020, \aap, 636, A76

\bibitem[{{Spada} {et~al.}(2011){Spada}, {Lanzafame}, {Lanza}, {Messina}, \&
  {Collier Cameron}}]{2011MNRAS.416..447S}
{Spada}, F., {Lanzafame}, A.~C., {Lanza}, A.~F., {Messina}, S., \& {Collier
  Cameron}, A. 2011, \mnras, 416, 447

\bibitem[{{Spruit}(2002)}]{2002A&A...381..923S}
{Spruit}, H.~C. 2002, \aap, 381, 923

\bibitem[{{Stanton} \& {Murillo}(2016)}]{2016PhRvE..93d3203S}
{Stanton}, L.~G. \& {Murillo}, M.~S. 2016, \pre, 93, 043203

\bibitem[{{Thoul} {et~al.}(1994){Thoul}, {Bahcall}, \&
  {Loeb}}]{1994ApJ...421..828T}
{Thoul}, A.~A., {Bahcall}, J.~N., \& {Loeb}, A. 1994, \apj, 421, 828

\bibitem[{{Timmes} \& {Swesty}(2000)}]{Timmes2000}
{Timmes}, F.~X. \& {Swesty}, F.~D. 2000, \apjs, 126, 501

\bibitem[{{van Saders} {et~al.}(2016){van Saders}, {Ceillier}, {Metcalfe},
  {Silva Aguirre}, {Pinsonneault}, {Garc{\'\i}a}, {Mathur}, \&
  {Davies}}]{2016Natur.529..181V}
{van Saders}, J.~L., {Ceillier}, T., {Metcalfe}, T.~S., {et~al.} 2016, \nat,
  529, 181

\bibitem[{{van Saders} \& {Pinsonneault}(2013)}]{2013ApJ...776...67V}
{van Saders}, J.~L. \& {Pinsonneault}, M.~H. 2013, \apj, 776, 67

\bibitem[{{Vidotto} {et~al.}(2014){Vidotto}, {Gregory}, {Jardine}, {Donati},
  {Petit}, {Morin}, {Folsom}, {Bouvier}, {Cameron}, {Hussain}, {Marsden},
  {Waite}, {Fares}, {Jeffers}, \& {do Nascimento}}]{2014MNRAS.441.2361V}
{Vidotto}, A.~A., {Gregory}, S.~G., {Jardine}, M., {et~al.} 2014, \mnras, 441,
  2361

\bibitem[{{Weber} \& {Davis}(1967)}]{1967ApJ...148..217W}
{Weber}, E.~J. \& {Davis}, Leverett, J. 1967, \apj, 148, 217

\bibitem[{{Weiss} {et~al.}(2004){Weiss}, {Hillebrandt}, {Thomas}, \&
  {Ritter}}]{2004cgps.book.....W}
{Weiss}, A., {Hillebrandt}, W., {Thomas}, H.~C., \& {Ritter}, H. 2004, {Cox and
  Giuli's Principles of Stellar Structure}

\bibitem[{{Zahn} {et~al.}(2007){Zahn}, {Brun}, \&
  {Mathis}}]{2007A&A...474..145Z}
{Zahn}, J.~P., {Brun}, A.~S., \& {Mathis}, S. 2007, \aap, 474, 145

\bibitem[{{Zaire} {et~al.}(2022){Zaire}, {Jouve}, {Gastine}, {Donati}, {Morin},
  {Landin}, \& {Folsom}}]{2022MNRAS.517.3392Z}
{Zaire}, B., {Jouve}, L., {Gastine}, T., {et~al.} 2022, \mnras, 517, 3392

\end{thebibliography}

\end{document}